\documentclass{JHEP3} % 10pt is ignored!

\JHEPspecialurl{http://jhep.sissa.it/JOURNAL/JHEP3.tar.gz}

\usepackage{epsfig,multicol,bbm}
\usepackage{graphicx}
\newcommand{\lsim}{\mathrel{\mathop{\kern 0pt \rlap
  {\raise.2ex\hbox{$<$}}}
  \lower.9ex\hbox{\kern-.190em $\sim$}}}
\newcommand{\gsim}{\mathrel{\mathop{\kern 0pt \rlap
  {\raise.2ex\hbox{$>$}}}
  \lower.9ex\hbox{\kern-.190em $\sim$}}}

\newcommand{\be}{\begin{equation}}
\newcommand{\ee}{\end{equation}}
\newcommand{\beqarr}{\begin{eqnarray}}
\newcommand{\eeqarr}{\end{eqnarray}}

\def\ptmiss{\not\!\!{p_T}}
\def\beq{\begin{equation}}   %
\def\eeq{\end{equation}}   %
\def\ben{\begin{eqnarray}}   %
\def\een{\end{eqnarray}}   %

\title{Right-handed sneutrino dark matter in $\mathbf{U(1)'}$
seesaw models and its signatures at the LHC}

\author{Priyotosh Bandyopadhyay, Eung Jin Chun and Jong-Chul Park \\
Korea Institute for Advanced Study,
Hoegiro 85, Dongdaemun-gu, Seoul 130-722, Korea\\
Email: \email{priyotosh@kias.re.kr}, \email{ejchun@kias.re.kr}, \email{jcpark@kias.re.kr} }

\abstract{We suggest a real right-handed sneutrino, $\tilde N_1$, as a good
dark matter candidate in a supersymmetric $Z'$ model realizing the
seesaw mechanism. When the extra gaugino, $\tilde Z'$, is lighter
than $Z'$, the thermal freeze-out of the dark matter annihilation
to right-handed neutrinos, $ \tilde{N}_1 \tilde{N}_1 \to NN$,
through the $t$-channel $\tilde Z'$ exchange is shown to produce
the right dark matter density. It is essential to include the
decay and inverse decay of $N$ in this process, otherwise $N$
decouples too early and thus dark matter is overproduced. At
the LHC, the search for the seesaw mechanism can be made by
observing the signatures of $pp \to \tilde Z' \tilde Z' \to NN + \ptmiss$
 as $\tilde Z'$ can be copiously
produced from the cascade decays of gluinos/squarks, which is
complementary to the search of $pp\to Z' \to NN$. This may also
open up a promising new channel of finding the Higgs boson from
the displaced $N$ decay.
 }

\begin{document}

%\maketitle  IS IGNORED %%%%%%%%%%%

\section{Introduction}

Supersymmetry at the TeV scale has a nice feature that it provides
a good dark matter (DM) candidate, the lightest supersymmetric
particle (LSP) which is stable when R-parity is imposed for the
proton stability. A typical LSP in the supersymmetric Standard
Model is a neutral gaugino with a small mixture of Higgsinos whose
relic density is naturally in the right range due to its weak
interaction property \cite{Jungman95}. The supersymmetric Standard
Model may be extended to include a new sector in which the LSP
resides. The observed neutrino masses and mixing require such an
extension in various ways \cite{Seesaw}. A standard example is to
introduce  right-handed neutrinos which are singlets of the
Standard Model (SM) gauge group and have only  Yukawa interactions
with active neutrinos. Then, a right-handed sneutrino,
which is the scalar superpartner of a right-handed neutrino,
can be the LSP and thus a dark matter candidate.
However, the cosmic production
mechanism of the right-handed sneutrino dark matter can be
different from the usual neutralino.

If neutrinos are Dirac particles without invoking the seesaw
mechanism, the required neutrino Yukawa coupling is of order
$10^{-13}$, and thus the right-handed sneutrino can never be thermalized.
However, thermal regeneration through the tiny Dirac neutrino
Yukawa coupling can be effective
to produce a sizable abundance of the right-handed sneutrino \cite{Asaka05}.
If the seesaw mechanism  is realized at the TeV scale, one has the
neutrino Yukawa coupling of order $10^{-7}$ which, then, leads to the
overproduction of the right-handed sneutrino LSP. To avoid this, there must
be a thermalization process through which its relic density can
be suppressed appropriately.  One of the
earliest proposals for such a thermalization is to assume an
unconventionally large $A$ term leading to a sizable mixing
between the left-handed and right-handed sneutrinos
\cite{Arkani00}. More recently, the non-minimal supersymmetric
Standard Model was considered to allow a coupling of the right-handed neutrino with
an extra singlet field introduced in the Higgs sector \cite{Nmssm}. Note also that
one can invoke the inverse seesaw mechanism which allows for a
possibility of order-one neutrino Yukawa couplings \cite{Inverse}.

%\medskip

In this paper, we propose a new way of generating thermal
freeze-out density of the right-handed sneutrino dark matter in
$U(1)'$ seesaw models which may arise from a grand unification
group \cite{Langacker08}.  In the case of Dirac neutrino models
with no lepton number violation, the right thermal abundance was
shown to be generated through the $Z'$ resonance enhancement of
the right-handed sneutrino annihilation \cite{Lee07}. However,
this feature is not shared with $U(1)'$ models realizing the
seesaw mechanism in which (complex) right-handed sneutrino
components are split into two real scalar fields, an LSP and a
heavier state, with a large mass difference driven by the lepton
number violating soft (Majorana) mass term and they couple
non-diagonally to the $Z'$ gauge boson. Thus, the $Z'$ interaction
is typically too weak to prevent the overproduction of the
right-handed sneutrino LSP.

It will be shown that the right-handed sneutrino annihilation to
right-handed neutrinos through the $t$-channel $U(1)'$ gaugino,
$\tilde{Z}'$, exchange can be efficient enough to produce the
observed dark matter density. In this process, the decays and
inverse decays of the right-handed neutrinos through the small Yukawa
coupling play an important role in keeping the right-handed
neutrinos in thermal equilibrium and thus controlling the dark
matter relic density, which is a distinguishable feature of the
thermal history of the right-handed sneutrino dark matter compared
with the conventional neutralino LSP dark matter. From the
numerical analysis of the Boltzmann equations, we will show how
the dark matter freeze-out density arises depending particularly
on the $\tilde{Z}'$ mass and the effective neutrino mass
corresponding to the dark matter sneutrino Yukawa coupling.

Searching for an extra gauge boson $Z'$ through dilepton
resonances is one of the key issues in collider experiments
\cite{lhcZp}. A novel feature of $U(1)'$ seesaw models is that
right-handed neutrinos can be produced directly from the  $Z'$
decay and their subsequent decay leaves a clean signature of
same-sign dileptons appearing at sizable displaced vertices
depending on the corresponding neutrino mass
\cite{Keung83,Langacker84}. Observing these signals at the LHC
will indicate the Majorana nature of neutrinos confirming the
seesaw mechanism \cite{lhcN}. Furthermore, it provides an
interesting possibility to discover the Higgs boson through
displaced $b$-jets \cite{Bandyo10} coming from the right-handed
neutrino decay. The supersymmetric model with the right-handed
sneutrino LSP enjoys these features with the production  of
$\tilde{Z}'$ and its decay to the right-handed neutrino and
sneutrino dark matter.  Independently of the $Z'$ production,
$\tilde{Z}'$ can be produced copiously from the cascade
decays of gluinos and squarks, and thus the search for the
presence of the Majorana right-handed neutrino can be made more
effectively. In particular, it is expected to have a large number
of events for the same-sign dilepton or four displaced $b$-jets
plus large missing energy from pair-produced $\tilde{Z}'$ if it is
the next LSP.

%\medskip

This paper is organized as follows.  In Section 2, we introduce a
$U(1)'$ seesaw model, and analyze the masses of $Z'$ and $\tilde Z'$
and interaction vertices relevant for our calculation. In Section
3, the thermal abundance of right-handed sneutrino dark matter is
calculated for various model parameters such as the masses of $\tilde Z'$ and the right-handed (s)neutrino, the neutrino Yukawa coupling, and the $U(1)'$ gauge coupling. In Section 4, LHC signatures of the model are discussed focusing on the pair production of the right-handed neutrinos from the decay of $Z'$ and $\tilde Z'$, which can provide a test for the seesaw mechanism and a new channel for the Higgs production.  We conclude in Section 5.

\section{A supersymmetric $\mathbf{U(1)'}$ seesaw model}

Among various possibilities of an extra gauge symmetry $U(1)'$ and
the presence of the associated right-handed neutrinos
\cite{Langacker08}, we will take the $U(1)_\chi$ model for our
explicit analysis. The particle content of our $U(1)_\chi$ model
is as follows:
\begin{equation}\label{chargeqn}
\begin{array}{c|ccc|cc|c|cc}
SU(5) & 10_F & \bar{5}_F & 1 (N) & 5_H & \bar{5}_H & 1 (X) & 1
(S_1) & 1 (S_2) \cr \hline 2 \sqrt{10} Q' & -1 & 3 & -5 & 2 & -2 &
0 & 10 & -10 \cr
\end{array}
\end{equation}
where $SU(5)$ representations and $U(1)'$ charges of the SM
fermions ($10_F, \bar{5}_F$),  Higgs bosons ($5_H,
\bar{5}_H$), and additional singlet fields ($N, S, S'_{1,2}$) are
shown. Here $N$ denotes the right-handed neutrino, $X$ is an
additional singlet field fit into the 27 representation of $E_6$,
and we introduced more singlets $S_{1,2}$, vector-like under
$U(1)'$, to break $U(1)'$ and generate the Majorana mass term of
$N$. Note that the right-handed neutrinos carry the largest charge under
$U(1)_\chi$ and thus the corresponding $Z'$ decays dominantly to
right-handed neutrinos. Furthermore, the additional singlet field
$X$ is neutral under $U(1)_\chi$ so that it can be used to
generate a mass to the $U(1)'$ Higgsinos as will be discussed
below. The gauge invariant superpotential in the seesaw sector is given by
\begin{equation}\label{wpot}
W_{seesaw} = y_{ij} L_i H_u N_j + {\lambda_{N_i}\over2} S_1 N_i
N_i\;,
\end{equation}
where $L_i$ and $H_u$ denote the lepton and Higgs doublet
superfields, respectively. After the $U(1)'$ breaking by the
vacuum expectation value $\langle S_1 \rangle$, the right-handed
neutrinos obtain the mass $m_{N_i}=\lambda_{N_i} \langle S_1
\rangle$ and induce the seesaw mass for the light neutrinos:
\begin{equation}
 \widetilde{m}^\nu_{ij} = - y_{ik} y_{jk}{ \langle H_u^0\rangle^2 \over
 m_{N_k} }\,.
\end{equation}
To generate the atmospheric neutrino mass $\widetilde{m}_\nu=0.05$
eV with the right-handed neutrino mass scale of $m_N\sim 100$ GeV,
we need the Yukawa coupling  of $y_\nu \sim 4\times 10^{-7}$.  For
the breaking of $U(1)'$, one can consider the following schemes.
The first option is to introduce the $\mu'$ term \cite{Khalil07}:
\begin{equation}
 W' = \mu' S_1 S_2 \,,
\end{equation}
and arrange radiative $U(1)'$ breaking with a large Yukawa
coupling, say $\lambda_{N_3}$, mimicking the electroweak breaking
in the Minimal Supersymmetric Standard Model (MSSM). Another possibility
is to consider
\begin{equation}
W'=  \lambda X S_1 S_2 + {\kappa\over3} X^3
\end{equation}
as in the non-minimal Higgs sector \cite{Ellwanger09}. The
corresponding scalar potential is given by
\begin{equation}
V=V_{\rm susy} + V_{\rm soft} + V_D\;,
\end{equation}
where
\begin{eqnarray}
V_{\rm susy} &=& \sum_{\phi=X,S_1,S_2,N_i,H_u,L_i} \left|
{\partial W \over \partial \phi}\right|^2\;, \\
V_{\rm soft} &=& \left[ y_{\nu}A_{\nu} \tilde{L}H_u\tilde{N}
+ \frac{\lambda_{N}}{2} A_{N} S_1\tilde{N}\tilde{N} + \lambda A_S X S_1 S_2
+ {\kappa \over 3} A_\kappa X^3 + h.c.\right] \nonumber\\
&& + m_X^2 |X|^2 + m_S^2 \left[|S_1|^2 + |S_2|^2\right]
+ m_{\widetilde{N}}^2 |\widetilde{N}|^2 \;, \label{soft}\\
V_D &=& {g^{\prime 2}\over2} \left[ Q'_{S_1} |S_1|^2 +Q'_{S_2} |S_2|^2
+Q'_N |\widetilde{N}|^2
+ \cdots \right]^2\;.
\end{eqnarray}
In this case, the $\mu'$ term is generated through  the vacuum
expectation value of $X$: $\mu'= \lambda \langle X \rangle$ which,
in the leading term, is  given by
\begin{equation}
\mu' \sim {\lambda \over 4 \kappa} \left(
 A_\kappa + \sqrt{A_\kappa^2-8 m_S^2}\right)\;,
\end{equation}
where $A_\kappa$ is the trilinear soft term and $m_S$ is the soft
mass of $S$ \cite{Ellwanger09}. The vacuum expectation values of
$S_{1,2}$ can be developed radiatively for $\lambda\sim 1$.

Given the vacuum expectation values $\langle S_{1,2}\rangle$, one
gets the $Z'$ gauge boson mass: $M_{Z'}^2= 2 g'^2 \sum_i
Q'^2_{S_i} \langle S_i\rangle^2$ where $g'$ denotes the $U(1)'$
gauge coupling. At the moment, the most stringent bound on the
$Z'$ mass comes from the electroweak precision data. For the
$U(1)_\chi$ model, we get $M_{Z'} > 1.14$ TeV for the reference
gauge coupling $g'=\sqrt{5/3} g_2 \tan \theta_W \approx 0.46$
\cite{Z'EWPD}, which gives $ \langle S_i \rangle > 800
\,\mbox{GeV}$ for $\langle S_1 \rangle = \langle S_2 \rangle$.
Note that the value of the gauge coupling $g'$ can be scaled down
depending on the particle content below the grand unification
scale. For our numerical analysis, we will take the above
reference value.

Important mass parameters for our analysis are the $U(1)'$ gaugino
mass and the dark matter mass. Defining $\tan\beta'= \langle S_2
\rangle/\langle S_1 \rangle$, the $U(1)'$ gaugino-Higgsino mass
matrix  in the basis of $[\tilde{Z}', \tilde{S}_1, \tilde{S}_2]$
is given by
\begin{equation}\label{massen}
 {\cal M} = \left[ \begin{array}{ccc}
          m_M &  M_{Z'} c_{\beta'} & - M_{Z'}s_{\beta'} \cr
         M_{Z'} c_{\beta'} & 0 & \mu' \cr
         - M_{Z'} s_{\beta'} & \mu' & 0 \cr \end{array} \right]\;,
\end{equation}
where $m_M$ is the soft supersymmetry breaking mass. In the limit
of $\mu' \gg m_M, M_{Z'}$, the lightest state has the mass
$m_{\tilde{Z}'}$ given by  $m_{\tilde{Z}'} \approx m_M + M_{Z'}^2
s_{2\beta'}/\mu'$. In the following, we will work in this limit.
As a light $\tilde{Z}'$ is favored for our dark matter relic
density, we will work in this limit taking $m_{\tilde{Z}'}$ as a
free parameter without bothering the specific values of
$\tan\beta'$ and $\mu'$. Among three right-handed neutrino
superfields, let us take the lightest one $N$ (we will suppress
the flavor index in the following discussions) whose scalar
component contains the LSP.  Denoting its supersymmetric mass by
$m_N$, the right-handed sneutrino $\tilde N$ has the mass terms:
\begin{equation}
V_{mass} = (m_N^2+m_{\tilde N}^2-{1\over4} M_{Z'}^2 c_{2\beta'})
|\tilde{N}|^2 - {1\over2} B_N m_N (\tilde N \tilde N
+ \tilde N^* \tilde N^*)\;,
\end{equation}
where $B_N m_N =- \lambda_{N} A_{N} \langle S_1 \rangle$. Due to
the lepton number violating (Majorana) mass term of $B_N m_N$
which is assumed to be positive, the real and imaginary components
of the sneutrino, $\tilde{N}=(\tilde N_1 + i \tilde
N_2)/\sqrt{2}$, get a mass splitting and their masses are given by
$m^2_{\tilde m_{1,2}}=
m_N^2+m_{\tilde N}^2-{1\over4} M_{Z'}^2 \mp B_N m_N$.
The real scalar field $\tilde N_1$
is taken to be the LSP dark matter.

\subsection{Couplings of $N$ and $\tilde{N}_1$ with Higgses}

From the superpotential in Eq.~(\ref{wpot}),
simplified here as $W_{seesaw}=y_\nu L H_u N + {1\over2} m_N NN$,  and the soft terms in
Eq.~(\ref{soft}), one can find the  couplings of $N$ and $\tilde{N}$
involving leptons/sleptons and the Higgs bosons.
For this, one uses the usual transformation to the Higgs mass eigenstates as follows:
\begin{eqnarray}\label{Higgsen}
H^0_d = \frac{1}{\sqrt2}\left( v_d  + h^0_d + i a^0_d\right)\;, &&
H^0_u = \frac{1}{\sqrt2}\left(v_u + h^0_u + i a^0_u \right)\;, \nonumber\\
h^0_d = \cos{\alpha}\,H -\sin{\alpha}\,h\;, &&
h^0_u = \sin{\alpha}\,H + \cos{\alpha}\,h\;, \nonumber\\
a^0_d = \sin{\beta}\, A - \cos{\beta}\, G^0\;, &&
a^0_u = \cos{\beta}\, A + \sin{\beta}\, G^0\;, \nonumber\\
H^\pm_d = \sin{\beta}\, H^\pm - \cos{\beta}\, G^\pm\;, &&
H^\pm_u = \cos{\beta}\, H^\pm + \sin{\beta}\, G^\pm\;,
\end{eqnarray}
where $\tan{\beta}=v_u/v_d$ and
$\tan{2\alpha}=\tan{2\beta}(\frac{m^2_A+M^2_Z}{m^2_A-M^2_Z})$.

Then the right-handed neutrino $N$ has the coupling with the left-handed neutrino $\nu$ and charged lepton $l$:
\begin{equation} \label{NHcoupling}
 {\cal L}_{NL} = {y_\nu \over \sqrt{2}} \,  N \nu  \left( h \cos\alpha + H \sin\alpha +i A \cos\beta\right)
 - y_\nu \, N l  H^+ \cos\beta + h.c.
\end{equation}
in the Weyl fermion notation. Due to the vacuum expectation values of the Higgs bosons, there arises mixing between $N$ and $\nu$ and the corresponding mixing angle is given by
\begin{eqnarray}\label{mixingf}
\theta_{N\nu} \approx \frac{y_{\nu}v_u}{\sqrt2 m_{N}}\;.
\end{eqnarray}
This mixing induces the $N$--$\nu$--$Z$ and $N$--$l$--$W^+$ interaction leading to the $N$ decay to the usual gauge bosons.
For completeness, let us write down the scalar interaction of $\tilde N=(\tilde N_1 + i \tilde N_2)/\sqrt{2}$:
\begin{eqnarray}
V_{\tilde N \tilde L} &=& y_\nu m_N \tilde \nu H^0_u \tilde N^*
 + y_\nu \tilde \nu ( A_L H^0_u - \mu H_d^{0*}) \tilde N \\
 && - y_\nu m_N \tilde l H^+_u \tilde N^*
 - y_\nu \tilde l ( A_L H^+_u + \mu H_d^{+}) \tilde N  + h.c.\,,\nonumber
\end{eqnarray}
from which one can read off the couplings in terms of the mass eigenstates.
The above scalar interaction leads also to mixing between $\tilde N$ and $\tilde \nu$ whose mixing angle is given by
\begin{eqnarray}\label{mixing}
\theta_{\tilde{N}_{1,2}\tilde{\nu}_{R,I}}
= y_\nu \frac{(m_N \pm A_L \mp \mu \cot{\beta})v_u}
{\sqrt2(m^2_{\tilde{N}_{1,2}}-m^2_{\tilde{\nu}_{R,I}})}\;.
\end{eqnarray}

\subsection{Decay of the right-handed neutrino N}\label{Ndecays}

From the couplings (\ref{NHcoupling}) and the mixing (\ref{mixingf}), various
channels open up as the decay modes of the right-handed neutrino, N.
The corresponding decay widths are given as follows:
\begin{eqnarray}\label{ndcy}
\Gamma(N\to\nu h) &=& \Gamma \left( N\to\bar{\nu} h \right)
= \frac{1}{8}\frac{y^2_{\nu}m_N}{8\pi} \left(1-\frac{m^2_h}{m_N^2}\right)^2\cos^2{\alpha}\;,\\\nonumber
\Gamma(N\to\nu H) &=& \Gamma \left( N\to\bar{\nu} H \right)
= \frac{1}{8}\frac{y^2_{\nu}m_N}{8\pi} \left(1-\frac{m^2_H}{m_N^2}\right)^2\sin^2{\alpha}\;,\\\nonumber
\Gamma(N\to\nu A) &=& \Gamma \left( N\to\bar{\nu} A \right)
= \frac{1}{8}\frac{y^2_{\nu}m_N}{8\pi} \left(1-\frac{m^2_A}{m_N^2}\right)^2\cos^2{\beta}\;,\\\nonumber
\Gamma(N\to\l H^+) &=& \Gamma(N\to\bar{\l} H^-)
= \frac{1}{4}\frac{y^2_{\nu}m_N}{8\pi} \left(1-\frac{m^2_{H^\pm}}{m_N^2}\right)^2\cos^2{\beta}\;,\\\nonumber
\Gamma(N\to\nu Z) &=& \Gamma (N\to\bar{\nu} Z)
= \frac{1}{8}\frac{y^2_{\nu}m_N}{8\pi}
\left(1-\frac{M^2_Z}{m_N^2}\right)^2 \left(1+2\frac{M^2_Z}{m_N^2}\right)\sin^2{\beta}\;,\\\nonumber
\Gamma(N\to\l W^+) &=& \Gamma(N\to\bar{\l} W^-)
= \frac{1}{8}\frac{y^2_{\nu}m_N}{8\pi} \left(1-\frac{M_W^2}{m_N^2}\right)^2
\left(1 + 2 \frac{M_W^2}{m_N^2}\right)\sin^2{\beta}\;.
\end{eqnarray}
To quantify these decays, we will use the notion of the effective neutrino mass defined by
\begin{equation}
 \tilde m_\nu \equiv {y_\nu^2 v_u^2 \over 2 m_N}
\end{equation}
which can be of order of the observed neutrino mass or smaller.

\subsection{Couplings  of $Z'$ and $\tilde Z'$}

The $U(1)'$ gauge boson and gaugino couplings in our model is
summarized in Table 1.  For the
Higgs--neutralino(chargino)--$\tilde Z'$ coupling, the
diagonalization matrices of the usual neutralinos and charginos
are defined by
\begin{eqnarray}
 \tilde H_{u}^0 = N_{4i} \lambda^0_i \,,&&
 \tilde H_{d}^0 = N_{3i} \lambda^0_i   \\
 \tilde H^+_u = V_{2i} \lambda^+_i \,,&&
 \tilde H^-_d = U_{2i} \lambda^-_i \nonumber
\end{eqnarray}
in the two component notation. Moreover, the four component mass
eigenstates are defined in terms of the two component fields as
\begin{equation}
\chi^0_i =
 \left(
{\begin{array}{cc}
\lambda^0_i \cr
\overline{\lambda}^0_i \cr
\end{array}}
 \right)\,, \quad
\chi^+_i =
 \left(
{\begin{array}{cc}
\lambda^+_i \cr
\overline{\lambda}^-_i \cr
\end{array}}
 \right)\,.
\end{equation}

\begin{table}[h]
\begin{center}
\renewcommand{\arraystretch}{1.4}
\begin{tabular}{||c|c||c|c||}
\hline
 $\bar{f}fZ^{\prime}$ &
 $ig^\prime (Q'_{f_L} P_L - Q'_{f^\dag_R} P_R)\gamma^{\mu}$ &
 $hAZ^{\prime }$ &
 $-ig^\prime Q'_{H_u}\sin(\alpha +\beta)q^{\mu}$
 \\
 \hline
 $\tilde{f}\tilde{f}^* Z^{\prime}$ &
 $ig^\prime Q'_f q^{\mu} $ &
 $HAZ^{\prime}$ &
 $ig^\prime Q'_{H_u}\cos(\alpha +\beta)q^{\mu}$
 \\
 \hline
 $\tilde{f}_L \bar{f} \tilde{Z}'$ &
 $-i\sqrt{2}g'Q'_{f_L}P_R$ &
 $H^+H^- Z^{\prime }$ &
 $ig^\prime Q'_{H_u} q^\mu$
 \\
 \hline
 $\tilde{f}_R \bar{f} \tilde{Z}'$ &
 $-i\sqrt{2}g' Q'_{f_R^\dag}P_L$ &
 $h Z Z^{\prime}$ &
 ${-i\sqrt{2}g^\prime M_Z} Q'_{H_u}\cos(\alpha +\beta) g^{\mu\nu}$
 \\
 \hline
 $\tilde{N}_1\tilde{N}_2Z^{\prime}$ &
 $ig^\prime Q'_N q^{\mu} $ &
 $H Z Z^{\prime}$ &
 ${-i\sqrt{2}g^\prime M_Z} Q'_{H_u}\sin(\alpha +\beta) g^{\mu\nu}$
  \\
 \hline
 $\bar{N}\tilde{N}_{1} \tilde{Z^{\prime}}$ &
 $-ig^\prime Q'_NP_R$ &
 $h \bar{\chi}^0_i \tilde Z' $ &
 $-i g'Q'_{H_u}(N_{4i} \cos\alpha + N_{3i}\sin\alpha) P_R$
  \\
  \hline
 $\bar{N}\tilde{N}_{2} \tilde{Z^{\prime}}$ &
 $g^\prime Q'_NP_R$ &
 $H\bar{\chi}^0_i \tilde Z' $ &
 $-i g'Q'_{H_u}(N_{4i} \sin\alpha - N_{3i}\cos\alpha) P_R$
  \\
 \hline
 ${S}_{1,2}S_{1,2}^* Z^{\prime}$ &
 $ig^\prime Q'_{S_{1,2}} q^{\mu}$ &
 $A \bar{\chi}^0_i \tilde Z' $ &
 $g'Q'_{H_u}(N_{4i} \cos\beta - N_{3i}\sin\beta) P_R$
  \\
 \hline
 $\bar{\tilde{S}}_{1,2}\tilde{S}_{1,2}Z^{\prime}$ &
 $ig^\prime Q'_{S_{1,2}}\gamma^{\mu}$&
 $H^+ \bar{\chi}^+_i \tilde Z' $ &
  $-i\sqrt{2} g'Q'_{H_u}(V_{2i}^* \cos\beta\, P_R - U_{2i} \sin\beta\, P_L)$
 \\
 \hline
 $S_{1,2} \bar{\tilde{S}}_{1,2}  \tilde{Z^{\prime}}$ &
 $-i\sqrt{2}g^\prime Q'_{S_{1,2}}P_R$ &
  &
     \\
 \hline

\end{tabular}

\caption{Couplings of $Z'$ and $\tilde Z'$ in the $U(1)_\chi$
model where $g'$ and $Q_X$ are the $U(1)_\chi$ gauge coupling and
charge of the field $X$, respectively, and  $q^\mu$ denotes the
4-momentum difference between two bosons in the
vertex.
}\label{couplzp}
\end{center}
\end{table}

\section{Relic density of the right-handed sneutrino dark matter}\label{DM}

In this section, we will calculate the thermal relic abundance of
the  right-handed sneutrino DM $\widetilde{N}_1$. Since the DM particle
 $\widetilde{N}_1$ cannot annihilate through the $Z'$ interaction,
one may consider its annihilation  to  right-handed
neutrinos through the $t$-channel  $\widetilde{Z}'$
exchange. In this case, the question is whether the right-handed neutrino $N$ remains
sufficiently in thermal equilibrium.
Consequently, it is very crucial to study the
interaction rate for reactions of $N$. As will be shown in detail,
the $Z'$ interaction of $N$ alone is too weak to
give the right dark matter relic density but the inclusion of
the decay and inverse decay of $N$ can significantly change the result.
For the following analyses, we will assume the numerical values
of $m_h = 115$ GeV,  $\tan\beta=10$,
and $g'=0.46$ unless otherwise stated.

\subsection{Freeze out of $\widetilde{N}_1$}

\begin{figure}
\begin{center}
\includegraphics[width=0.75\linewidth]{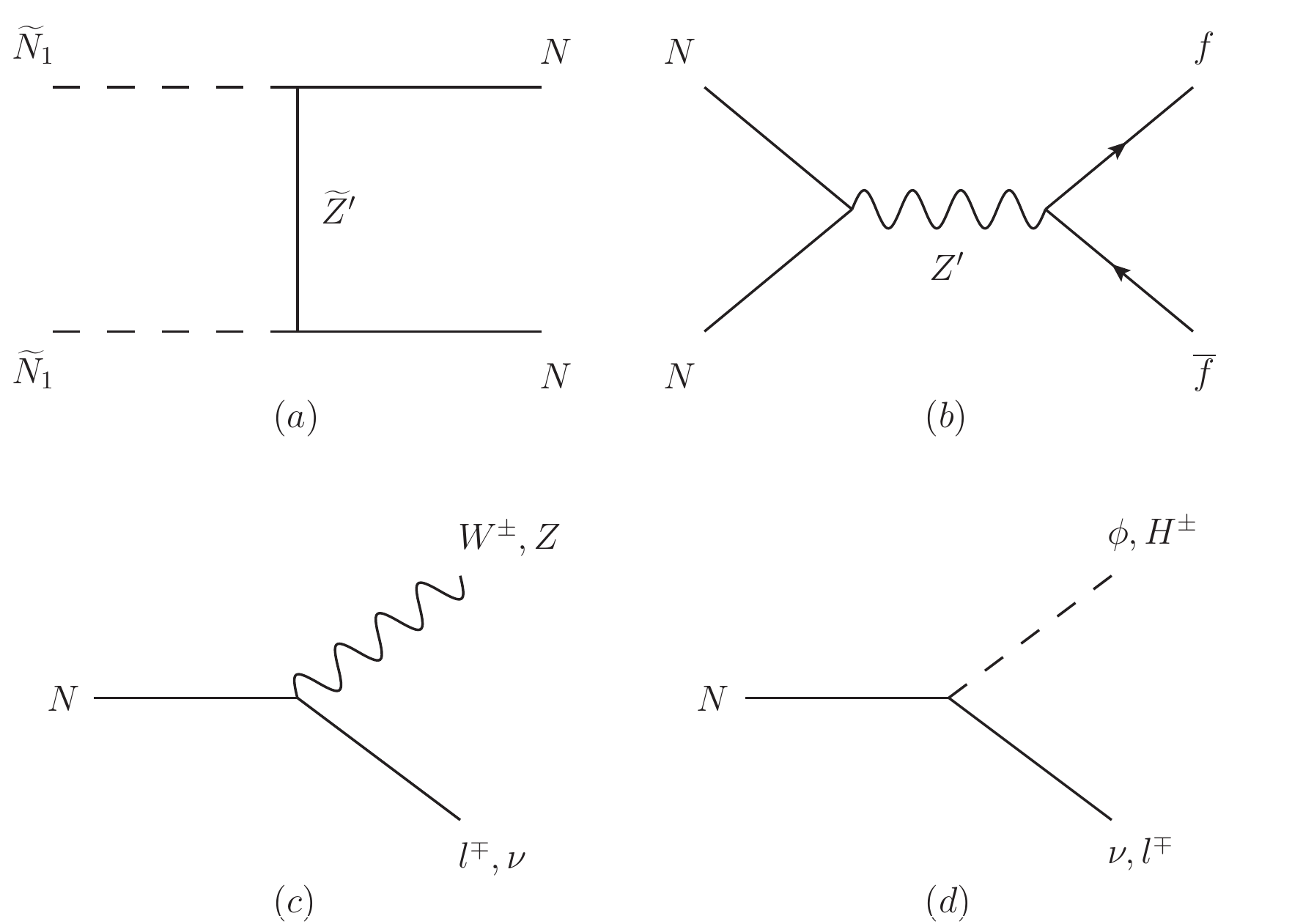}
\end{center}
\caption{Annihilation and decay channels of $\widetilde{N}_1$
and $N$. In panel (d), $\phi = h, H$ and $A$.}\label{Diagrams}
\end{figure}

The annihilation cross section of the DM candidate $\tilde N_1$
through the process,
 $\tilde N_1 \tilde N_1 \to NN$, can
be large enough to suppress arbitrarily its relic density
depending on mass parameters such as $m_{\widetilde{N}_1}, m_N$,
and $m_{\widetilde{Z}'}$ as far as $N$ is in thermal equilibrium
long enough. However, the interaction of $N$ is either through the
heavy gauge boson $Z'$ or the tiny neutrino Yukawa coupling
$y_\nu$ and thus might be too weak to keep $N$ in thermal
equilibrium during the process of  freeze-out of $\tilde N_1$.
Thus, to study the thermal history of  $\widetilde{N}_1$ through
the annihilation in Figure \ref{Diagrams}(a), one has to consider
also the evolution of $N$ determined by its annihilation (Figure
\ref{Diagrams} (b)) and decay (Figure \ref{Diagrams} (c), (d)).

%
%
%%%%%%%%%%%%%%%%%%%%%%%%%%%%%%%%%%%%%%%%%%%%%%%%%%%%%%%%%%%%%%%
%%%%%%%%%%%%%%%%%%%%%%%%%%%%%%%%%%%%%%%%%%%%%%%%%%%%%%%%%%%%%%%%
%
%

For the analysis of the relic abundance of the $\widetilde{N}_1$, let us set up
the following coupled Boltzmann equations for the
evolution of the number density $n_i$ of particle $i$:
\begin{eqnarray}
\frac{dn_{\widetilde{N}_1}}{dt} &=& -3Hn_{\widetilde{N}_1}
- \langle \sigma_{\widetilde{N}_1}v_{\widetilde{N}_1} \rangle
\left[ (n_{\widetilde{N}_1})^2 -
\left( \frac{g_{\widetilde{N}_1}}{g_N} n_N \right)^2 \right]\;,
\label{Boltxmann1}\\
\frac{dn_N}{dt} &=& -3Hn_N - \langle \sigma_Nv_N \rangle
\left[ (n_N)^2 - (n^{\rm eq}_N)^2 \right]
+ \langle \sigma_{\widetilde{N}_1}v_{\widetilde{N}_1}
\rangle \left[ (n_{\widetilde{N}_1})^2 -
\left( \frac{g_{\widetilde{N}_1}}{g_N} n_N \right)^2 \right]
\label{Boltxmann2}\nonumber\\
&&~~~~~~~~~~~ - \Gamma_N (n_N- n^{\rm eq}_N)\;,
\end{eqnarray}
where $H$ is the Hubble parameter, and $n^{\rm eq}_i$, $v_i$ and
$g_i$ are respectively the equilibrium number density, relative
velocity and number of internal degrees of freedom of particle
$i$. The first terms on the right-hand sides of
Eqs.~(\ref{Boltxmann1}) and (\ref{Boltxmann2}) account for the
dilution due to the expansion of the Universe. The second term on
the right-hand side of Eq.~(\ref{Boltxmann1}) describes the
forward and backward reactions of $\widetilde{N}_1\widetilde{N}_1$
annihilation to $NN$ through the $t$-channel $\widetilde{Z}'$
exchange (Figure~\ref{Diagrams} (a)). The second term on the
right-hand side of Eq.~(\ref{Boltxmann2}) refers to the forward
and backward reactions of $NN$ annihilation to the SM fermion pairs
$f\overline{f}$ through the $s$-channel $Z'$ exchange
(Figure~\ref{Diagrams} (b)). The third term represents the effects
of $\widetilde{N}_1\widetilde{N}_1$ annihilation to $NN$ through
the $t$-channel $\widetilde{Z}'$ exchange (Figure~\ref{Diagrams}
(a)). The last term describes the decays and inverse decays of the
right-handed neutrinos shown in
%
% $N \rightarrow W^\pm l^\mp / Z \nu /h\nu$
%\footnote{$N\to H^\pm \l^\mp / A\nu/ H\nu$ effects are minimal.}
Figure~\ref{Diagrams} (c), (d).

In solving the Boltzmann equations (\ref{Boltxmann1}) and
(\ref{Boltxmann2}), it is useful to introduce the variable $Y_i
\equiv n_i/s$ describing the actual number of particle $i$ per
comoving volume, where $s$ is the entropy density of the Universe.
Solving the coupled differential equations (\ref{Boltxmann1}) and
(\ref{Boltxmann2}), one can find $Y_i$ as a function of $x \equiv
m_{\widetilde{N}_1}/T$.

%\medskip

In order to see the effect of the decay  more clearly,
let us first present the solutions of the coupled Boltzmann equations
excluding the last
term on the right-hand side of Eq.~(\ref{Boltxmann2}) which
accounts for the decays and inverse decays of the right-handed
neutrino.
%
%
%%%%%%%%%%%%%%%%%%%%%%%%%%%%%%%%%%%%%%%%%%%%%%%%%%%%%%%%%%%%%%%
\begin{figure}
\begin{center}
\includegraphics[width=0.65\linewidth]{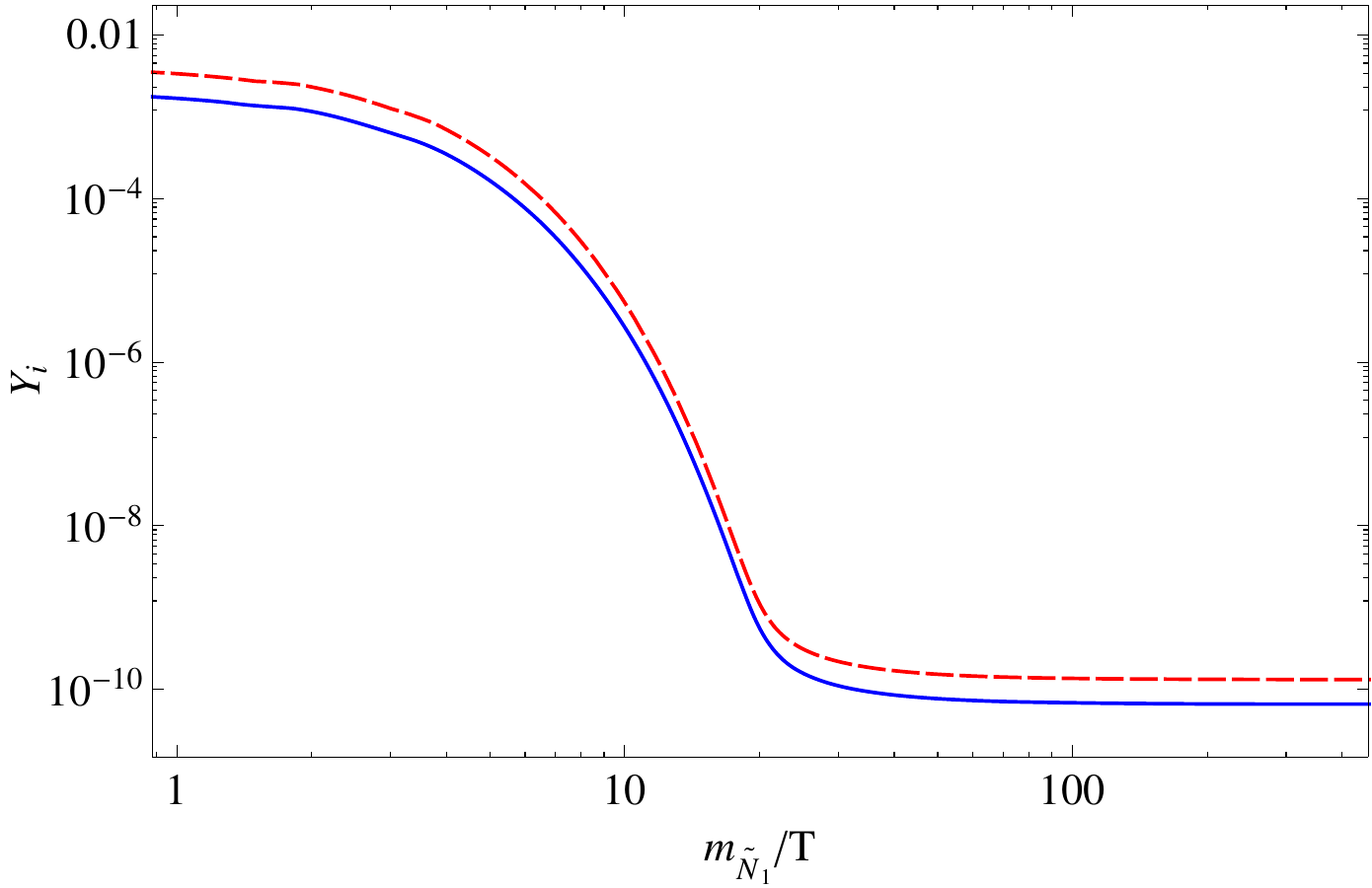}
\end{center}
\caption{The actual number of $\widetilde{N}_1$ and $N$ per
comoving volume without the decay effect term. Blue solid
and red dashed lines show $Y_{\widetilde{N}_1} \equiv
n_{\widetilde{N}_1}/s$ and $Y_N \equiv n_N/s$ respectively. The other
parameters are fixed as follows: $m_{\widetilde{N}_1} = 300$ GeV,
$m_N = 260$ GeV, $m_{\widetilde{Z}'} = 600$ GeV, and $M_{Z'} =
1200$ GeV.
}\label{fig:xf_NoDecay}
\end{figure}
%%%%%%%%%%%%%%%%%%%%%%%%%%%%%%%%%%%%%%%%%%%%%%%%%%%%%%%%%%%%%%%%
%
%
The result is shown in Figure~\ref{fig:xf_NoDecay}, where we use
the following mass parameters: $m_{\widetilde{N}_1} = 300$ GeV,
$m_N = 260$ GeV, $m_{\widetilde{Z}'} = 600$ GeV, and $M_{Z'} =
1200$ GeV. Since $\widetilde{N}_1$ can remain in thermal bath
through the interactions of $N$, the freeze out temperature of
$\widetilde{N}_1$ is determined by that of $N$, if the
interactions of $N$ are weaker than those of $\widetilde{N}_1$.
Therefore, without the decay effect term, $\widetilde{N}_1$ is
decoupled from thermal equilibrium when $N$ is decoupled as can be
seen from Figure~\ref{fig:xf_NoDecay}.

The right-handed neutrino mostly annihilates to SM
fermion-antifermion pairs $f\overline{f}$ through the $s$-channel
$Z'$ exchange, but the cross section for this process $NN
\rightarrow f\overline{f}$ is suppressed by large $Z'$ gauge boson
mass, $M_{Z'} > 1.14$ TeV~\cite{Z'EWPD}. Moreover, in the
zero-velocity limit, this annihilation cross section of
right-handed neutrinos is more suppressed since the annihilation
cross section for this process has no $s$-wave component. Thus,
the annihilation of the right-handed neutrino usually freezes out
earlier than that of the right-handed sneutrino DM candidate.
%
%
%%%%%%%%%%%%%%%%%%%%%%%%%%%%%%%%%%%%%%%%%%%%%%%%%%%%%%%%%%%%%%%
\begin{figure}
\begin{center}
\includegraphics[width=0.49\linewidth]{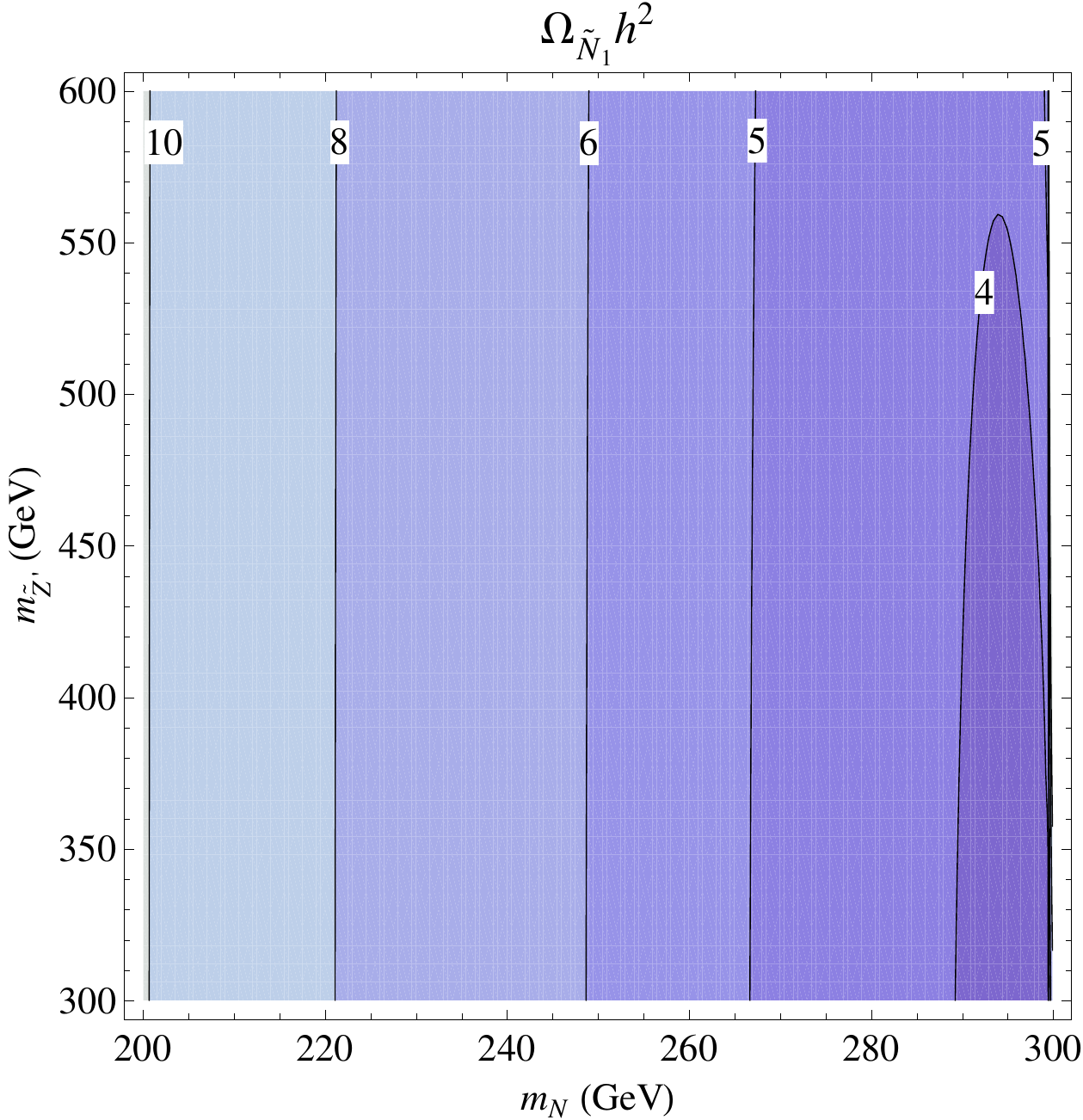}
\includegraphics[width=0.49\linewidth]{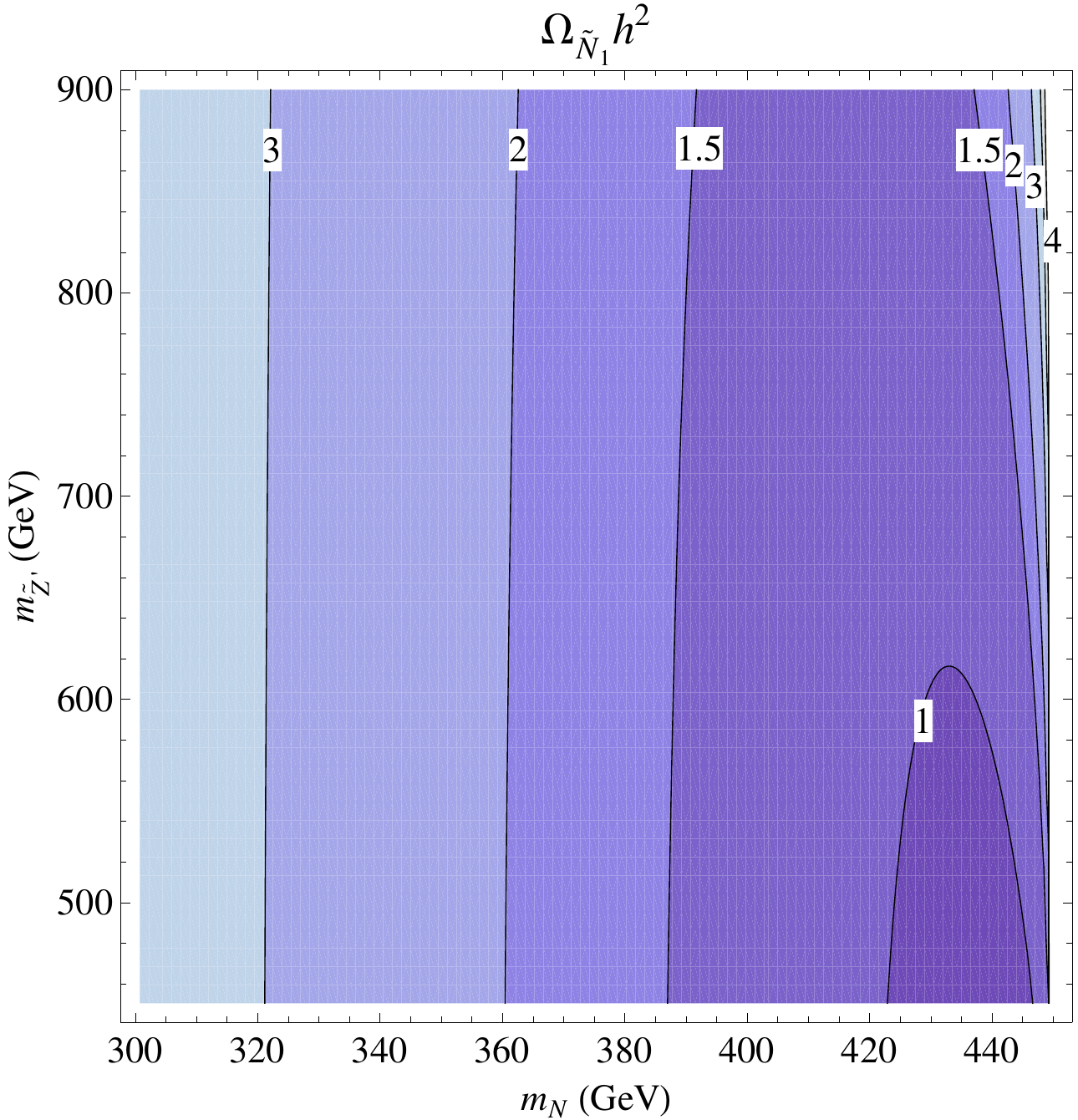}
\end{center}
\caption{Contour plots for the relic abundance of the right-handed
sneutrino dark matter $\widetilde{N}_1$ in the $m_N-m_{\widetilde{Z}'}$
plane when the decay effect of $N$ is excluded.
Each panel shows the cases
$m_{\widetilde{N}_1} = $ 300 and 450 GeV,
respectively. We fix the $Z'$ gauge boson mass as $M_{Z'} =
1200$ GeV.}\label{fig:mdm_NoDecay}
\end{figure}
%%%%%%%%%%%%%%%%%%%%%%%%%%%%%%%%%%%%%%%%%%%%%%%%%%%%%%%%%%%%%%%%
%
%
Consequently, the right-handed sneutrino $\widetilde{N}_1$ is
generically overproduced due to the early decoupling of $N$
 as shown in Figure
~\ref{fig:mdm_NoDecay}.

%\medskip

The decays and inverse decays of right-handed neutrinos through
the Yukawa coupling $y_\nu$ turn out to play a crucial role in
evading the previous overproduction problem.
%
%
%%%%%%%%%%%%%%%%%%%%%%%%%%%%%%%%%%%%%%%%%%%%%%%%%%%%%%%%%%%%%%%
\begin{figure}
\begin{center}
\includegraphics[width=0.495\linewidth]{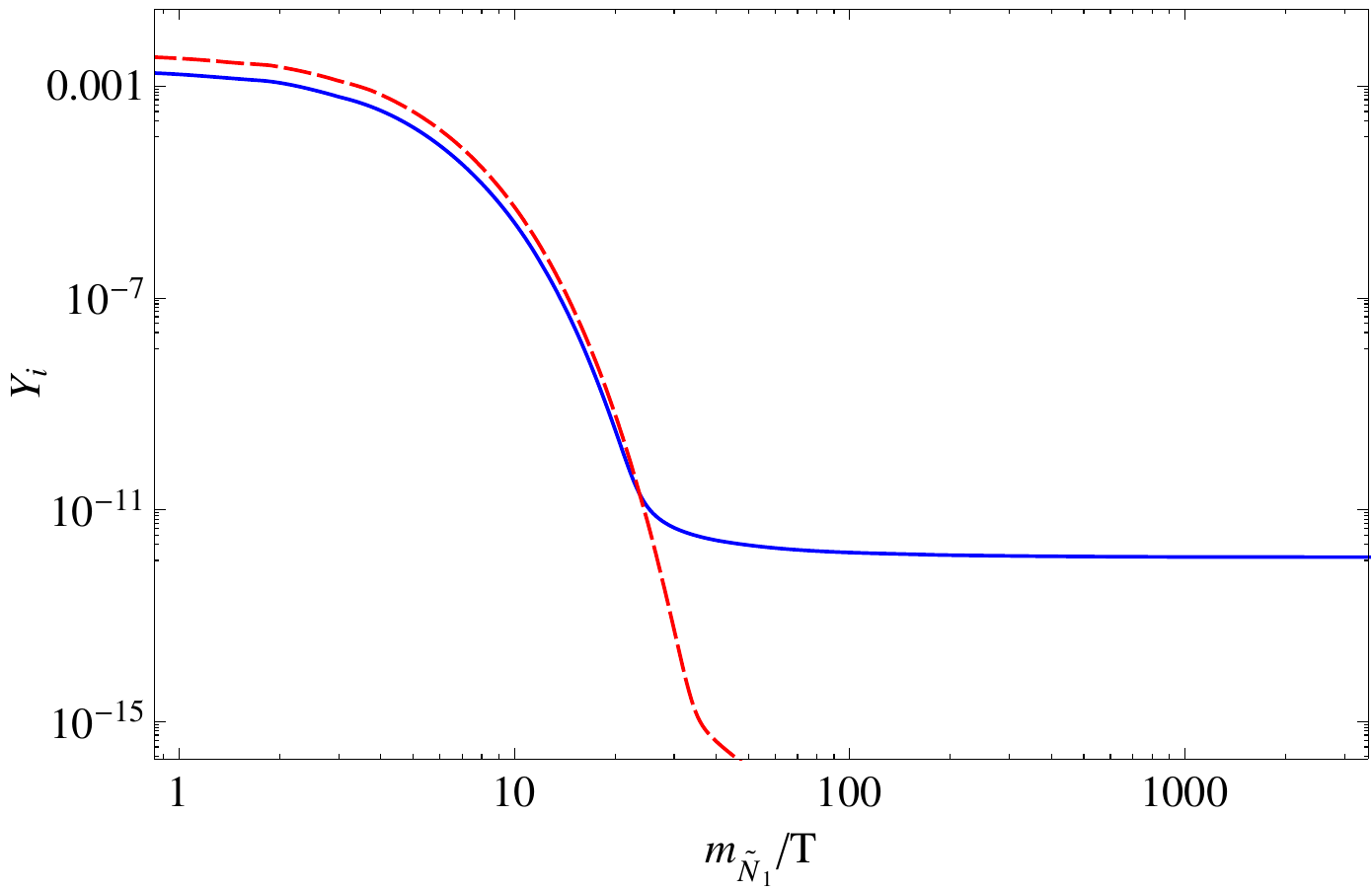}
\includegraphics[width=0.495\linewidth]{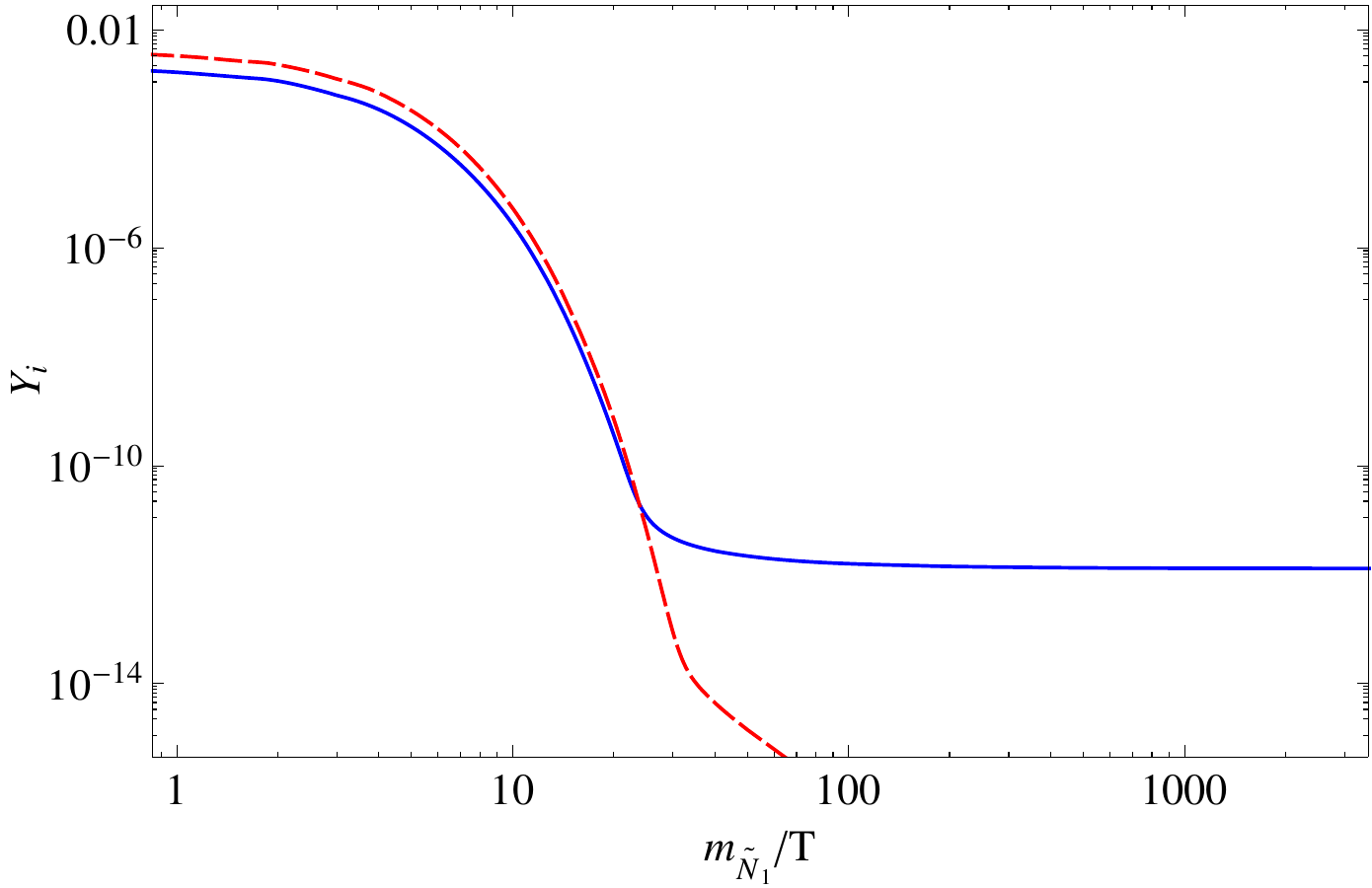}

\vspace*{0.3cm}

\includegraphics[width=0.495\linewidth]{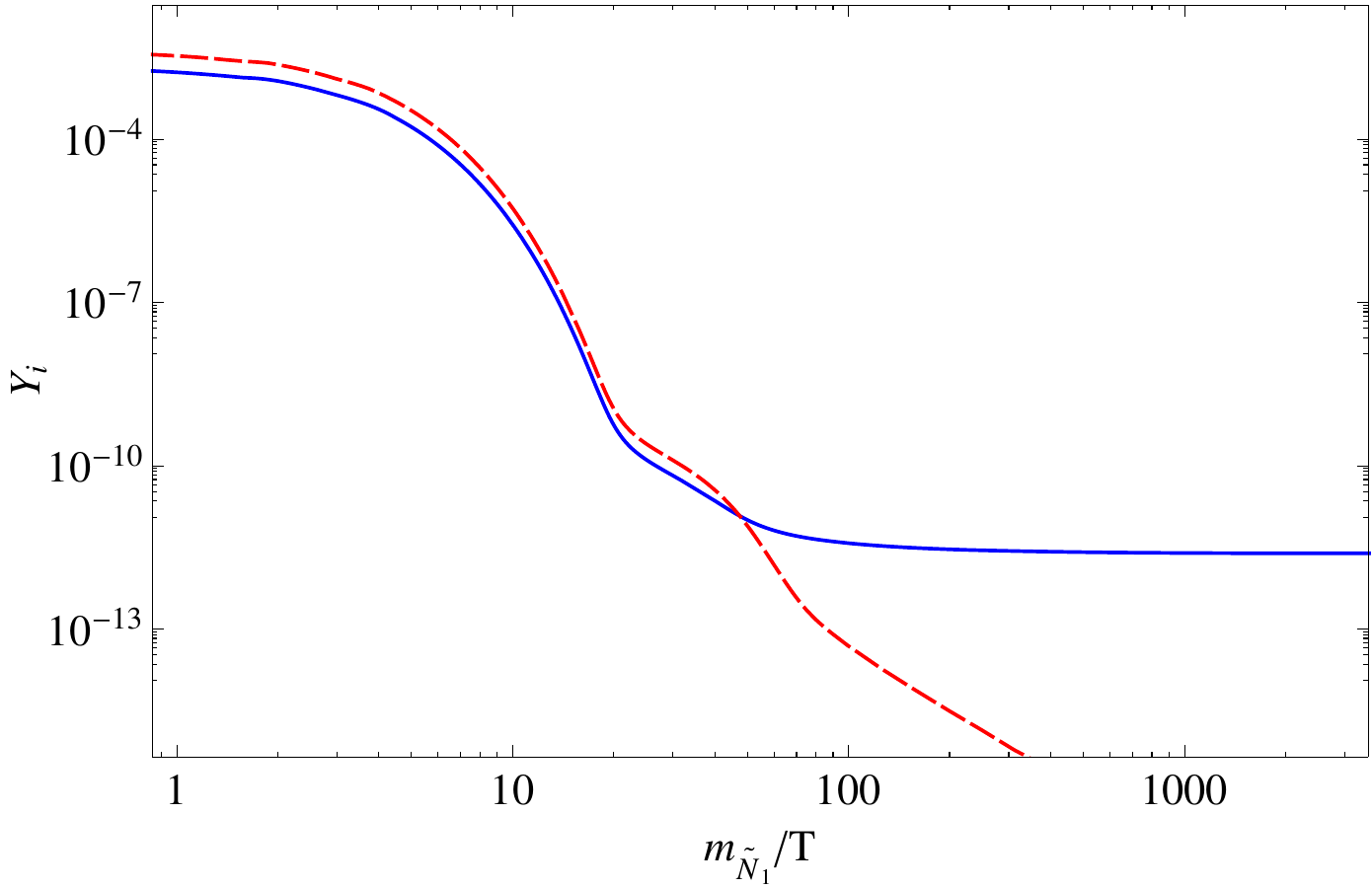}
\includegraphics[width=0.495\linewidth]{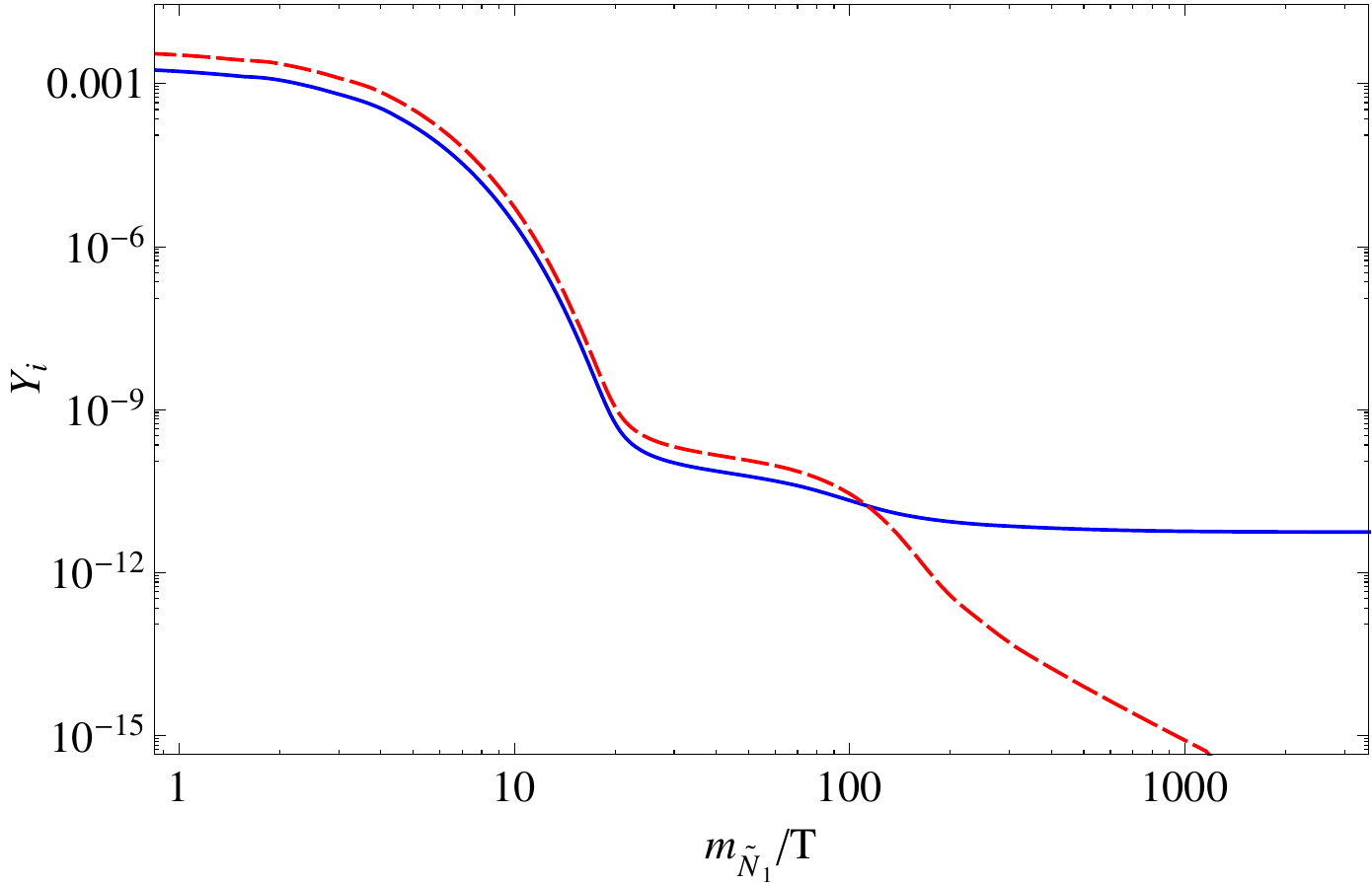}
\end{center}
\caption{The actual number of $\widetilde{N}_1$ and $N$ per
comoving volume. The panels correspond to $\widetilde{m}_\nu =
10^{-2}, 10^{-3}, 10^{-5}$ and $10^{-6}$ eV respectively from left
to right and top to bottom. Blue solid and red dashed lines show
$Y_{\widetilde{N}_1} \equiv n_{\widetilde{N}_1}/s$ and $Y_N \equiv
n_N/s$. The other parameters are the same as in the analysis of
Figure~\protect\ref{fig:xf_NoDecay}.}\label{fig:xf}
\end{figure}
%%%%%%%%%%%%%%%%%%%%%%%%%%%%%%%%%%%%%%%%%%%%%%%%%%%%%%%%%%%%%%%%
%
%
In Figure~\ref{fig:xf}, we present the evolution of the number density
of $\widetilde{N}_1$ and $N$ per comoving volume for four
representative values $\widetilde{m}_\nu = 10^{-2}, 10^{-3},
10^{-5}$, and $10^{-6}$ eV. In this analysis, we use the same
values for the other parameters as in the analysis of
Figure~\ref{fig:xf_NoDecay}. For larger values of light neutrino
mass ($\widetilde{m}_\nu \gtrsim 10^{-3}$ eV), $\textit{i.e.}$
larger Yukawa coupling $y_\nu$, the $N$ decay term of
Eq.~(\ref{Boltxmann2}) is strong enough to dominate other
interaction terms of $N$ before the annihilation effect of $N$
becomes weaker than the dilution effect due to the expansion of
the Universe. Therefore, the decay effect keeps $N$ in thermal
equilibrium for a longer time compared with the case that $N$ is
stable, and $N$ can continuously remain in thermal bath before
$\widetilde{N}_1$ is decoupled from thermal bath as can be seen
from the top two panels of Figure~\ref{fig:xf}. On the other hand,
for smaller light neutrino masses ($\widetilde{m}_\nu \lesssim
10^{-3}$ eV), the annihilation effect of $N$ becomes weaker than
the dilution effect before the $N$ decay effect dominates other
reactions. As a result, a retarded behavior appears in the
number density evolution of $\widetilde{N}_1$ and $N$ as can be
seen from the bottom two panels of Figure~\ref{fig:xf}. In
addition, since the decay modes of $N$ are governed by the Dirac
neutrino Yukawa coupling $y_\nu$, $\textit{i.e.}$ the effective neutrino
mass $\widetilde{m}_\nu$, the
result in Figure~\ref{fig:xf} converges to the one in
Figure~\ref{fig:xf_NoDecay} for smaller $\tilde m_\nu$.

\subsection{Dependence on $m_{\widetilde{N}_1}$}

In this and the following subsections, we will obtain the thermal
relic abundance of right-handed sneutrino
$\Omega_{\widetilde{N}_1}h^2$ given mass parameters including
$m_{\widetilde{N}_1}, m_N, \widetilde{m}_\nu, m_{\widetilde{Z}'}$,
$M_{Z'}$ and also the $U(1)'$ gauge coupling $g'$ by solving
the coupled Boltzmann equations (\ref{Boltxmann1}) and
(\ref{Boltxmann2}) including the decay term. In order to see the
dependence of the relic abundance  on each parameter,  we search
for the allowed regions, which satisfy the observed recent DM
relic density limit~\cite{WMAP7}, in the
$m_N$--$m_{\widetilde{Z}'}$ plane fixing the other parameters.

%
%
%%%%%%%%%%%%%%%%%%%%%%%%%%%%%%%%%%%%%%%%%%%%%%%%%%%%%%%%%%%%%%%
\begin{figure}
\begin{center}
\includegraphics[width=0.328\linewidth]{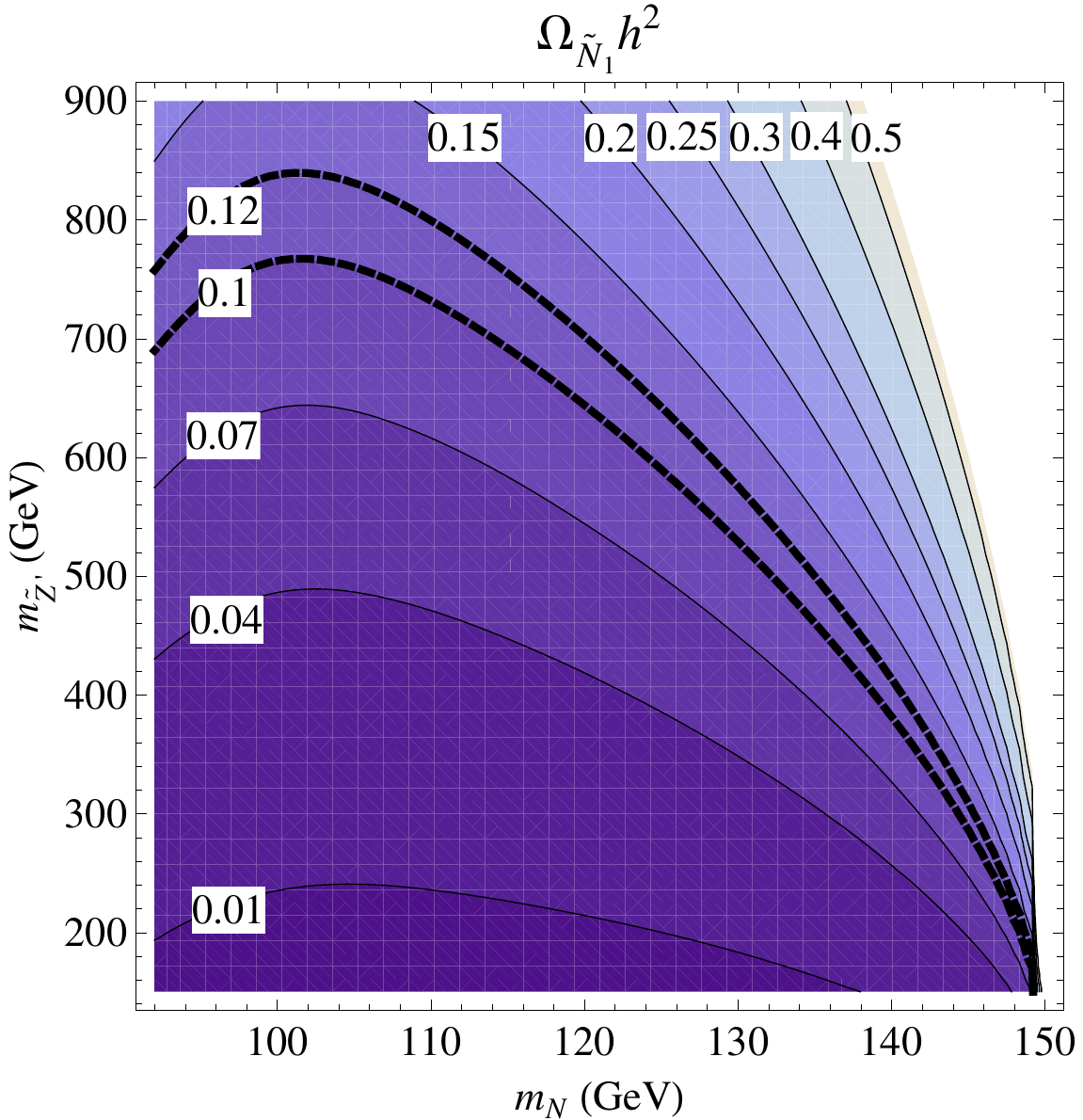}
\includegraphics[width=0.328\linewidth]{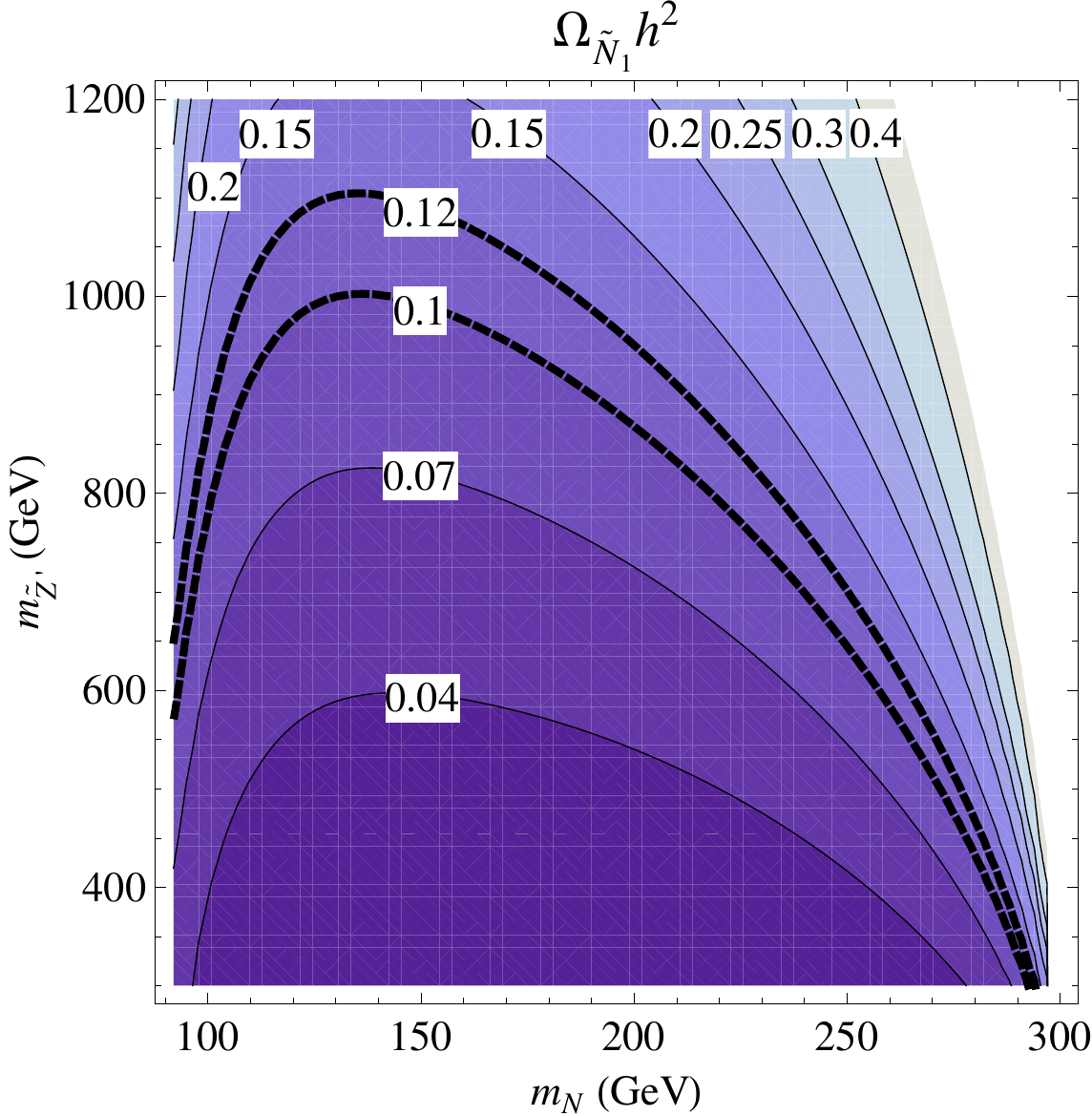}
\includegraphics[width=0.328\linewidth]{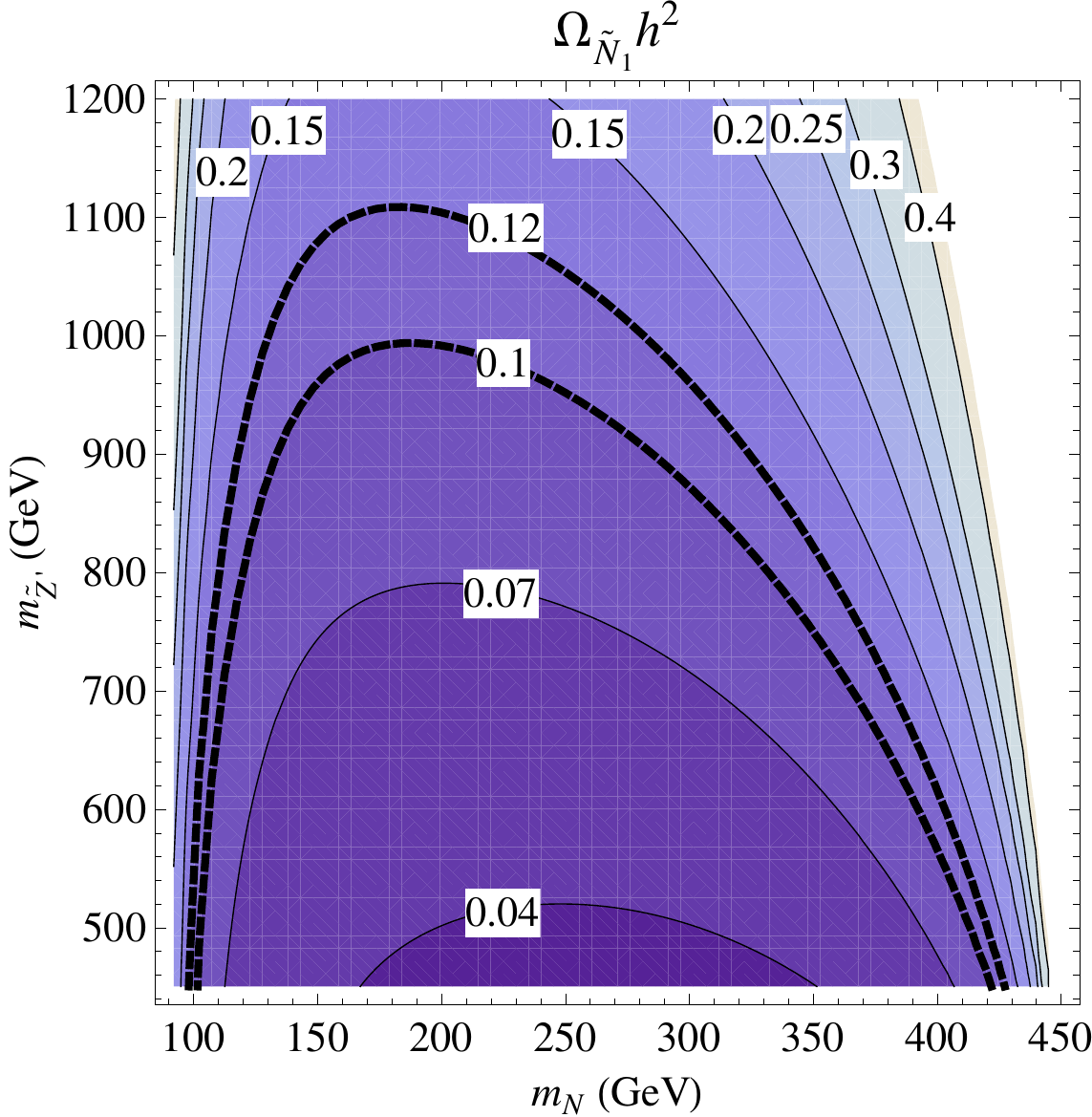}

\vspace*{0.2cm}

\includegraphics[width=0.328\linewidth]{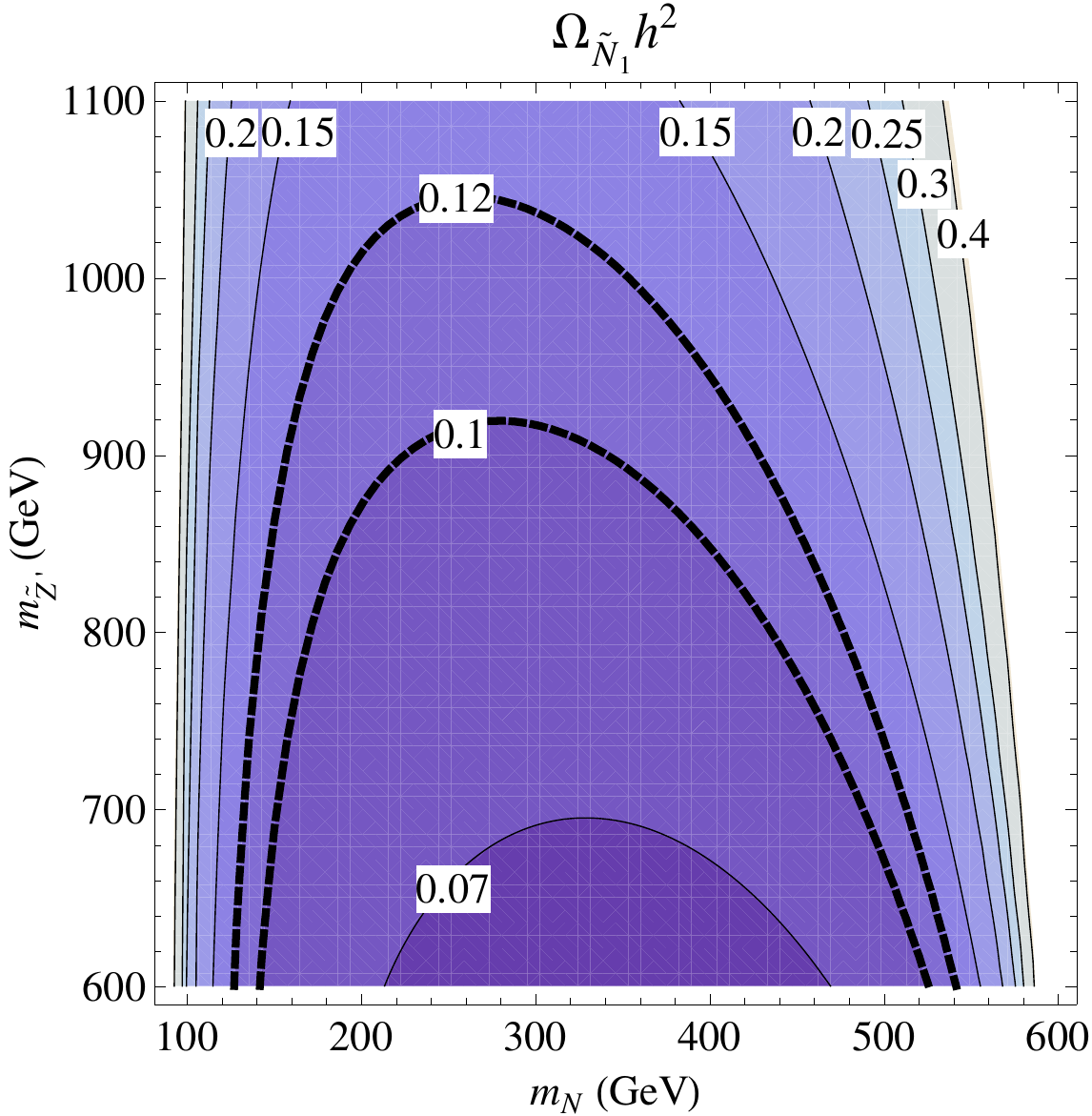}
\includegraphics[width=0.328\linewidth]{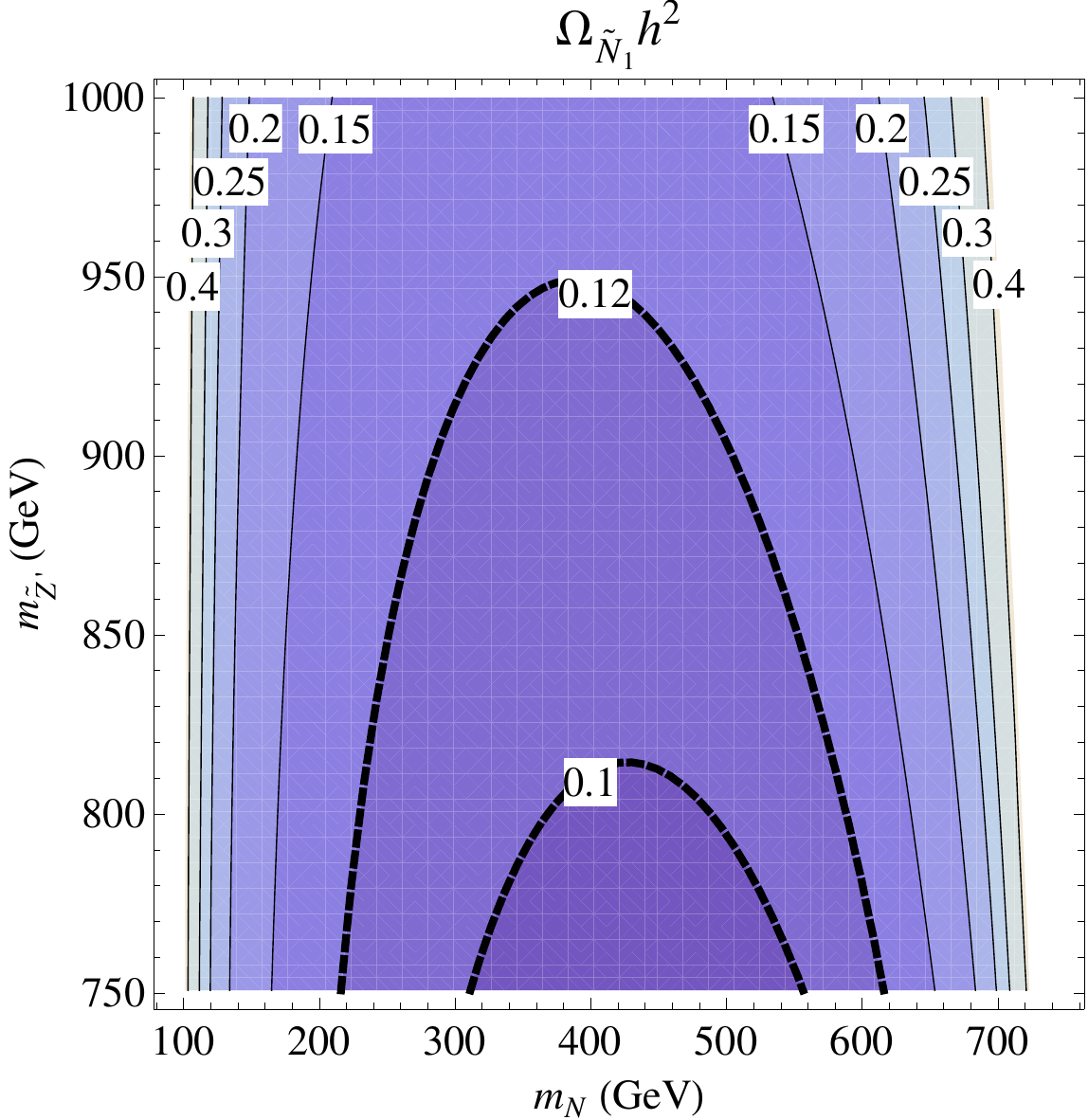}
\includegraphics[width=0.328\linewidth]{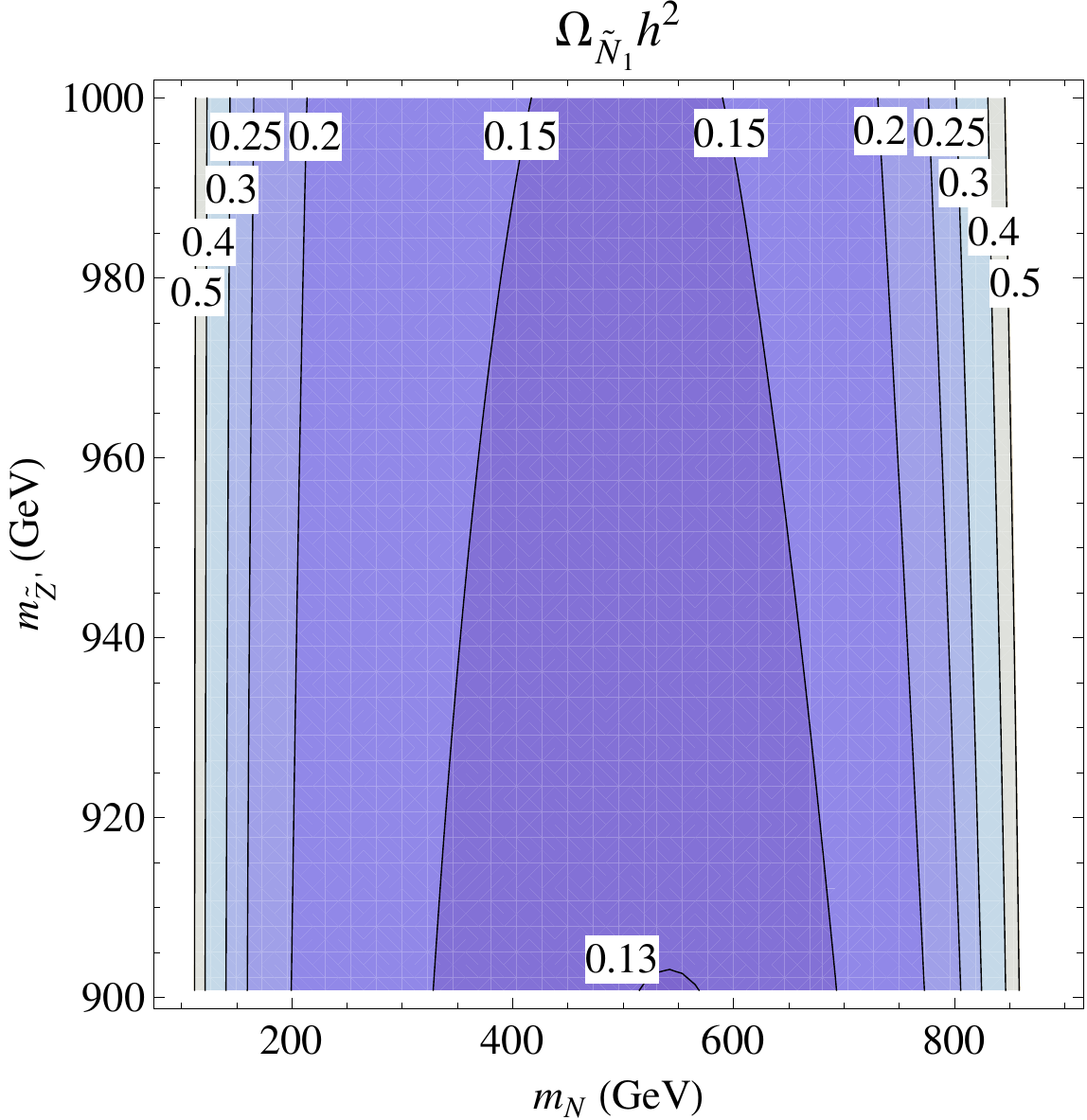}
\end{center}
\caption{Contour plots for the relic abundance of the right-handed
sneutrino dark matter $\widetilde{N}_1$ in the
$m_N-m_{\widetilde{Z}'}$ plane. The panels show the cases
$m_{\widetilde{N}_1} = $ 150, 300, 450, 600, 750 and 900 GeV
respectively from left to right and top to bottom. We fix the
other parameters as follows: $M_{Z'} = 1200$ GeV and
$\widetilde{m}_\nu = 10^{-3}$ eV. In each panel, the region
between two thick dashed lines is preferred by the recent result
on the DM relic density. }\label{fig:mdm}
\end{figure}
%%%%%%%%%%%%%%%%%%%%%%%%%%%%%%%%%%%%%%%%%%%%%%%%%%%%%%%%%%%%%%%%
%
%

In Figure~\ref{fig:mdm}, the $m_N$--$m_{\widetilde{Z}'}$ parameter
space
is explored for the six cases of the right-handed sneutrino dark matter mass:
 $m_{\widetilde{N}_1} = $ 150, 300,
450, 600, 750 and 900 GeV.
% in order to see the dependence of the DM
%relic density $\Omega_{\widetilde{N}_1}h^2$ on the dark matter
%mass $m_{\widetilde{N}_1}$.
Our numerical analysis is performed in
the case of $M_{Z'} = 1200$ GeV and $\widetilde{m}_\nu = 10^{-3}$
eV. The region between two thick dashed lines represents points
where $\Omega_{\widetilde{N}_1}h^2$ is consistent with the recent
WMAP result on the DM relic density~\cite{WMAP7}. As one can see
in Figure~\ref{fig:mdm}, the right-handed sneutrino mass less than
900 GeV, $m_{\widetilde{N}_1} < 900$ GeV, is allowed by the recent
WMAP result.

\subsection{Dependence on $\widetilde{m}_\nu$}

The right-handed sneutrino DM particle $\widetilde{N}_1$ can
remain in thermal equilibrium through the interactions of the
right-handed neutrino $N$ for which the inclusion of the decay effect is crucial as shown already.
In this subsection, we study this effect in more detail by changing the neutrino Yukawa coupling
$y_\nu$, that is, the  effective neutrino mass $\tilde m_\nu$.
%
%
%%%%%%%%%%%%%%%%%%%%%%%%%%%%%%%%%%%%%%%%%%%%%%%%%%%%%%%%%%%%%%%
\begin{figure}
\begin{center}
\includegraphics[width=0.493\linewidth]{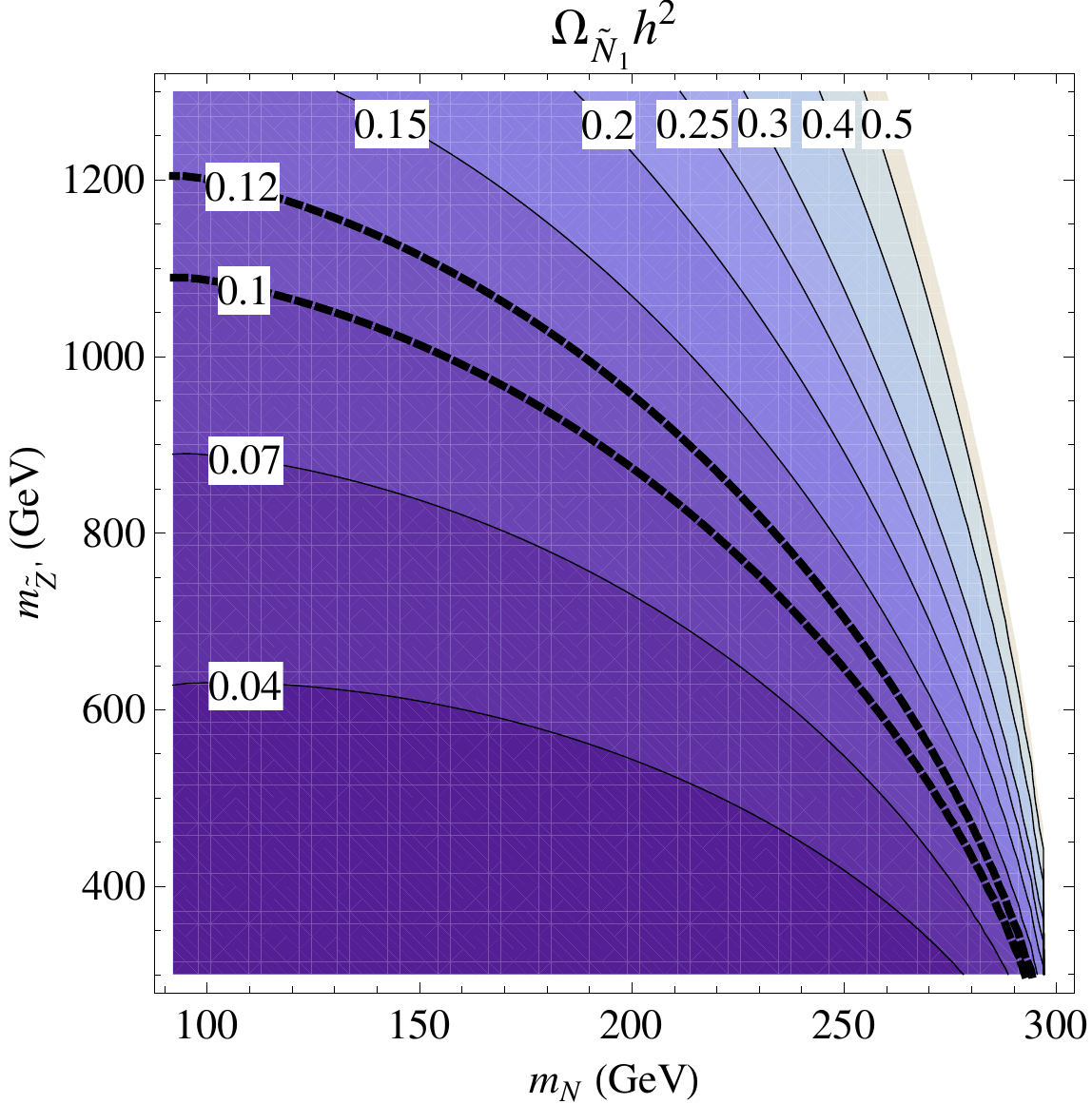}
\includegraphics[width=0.487\linewidth]{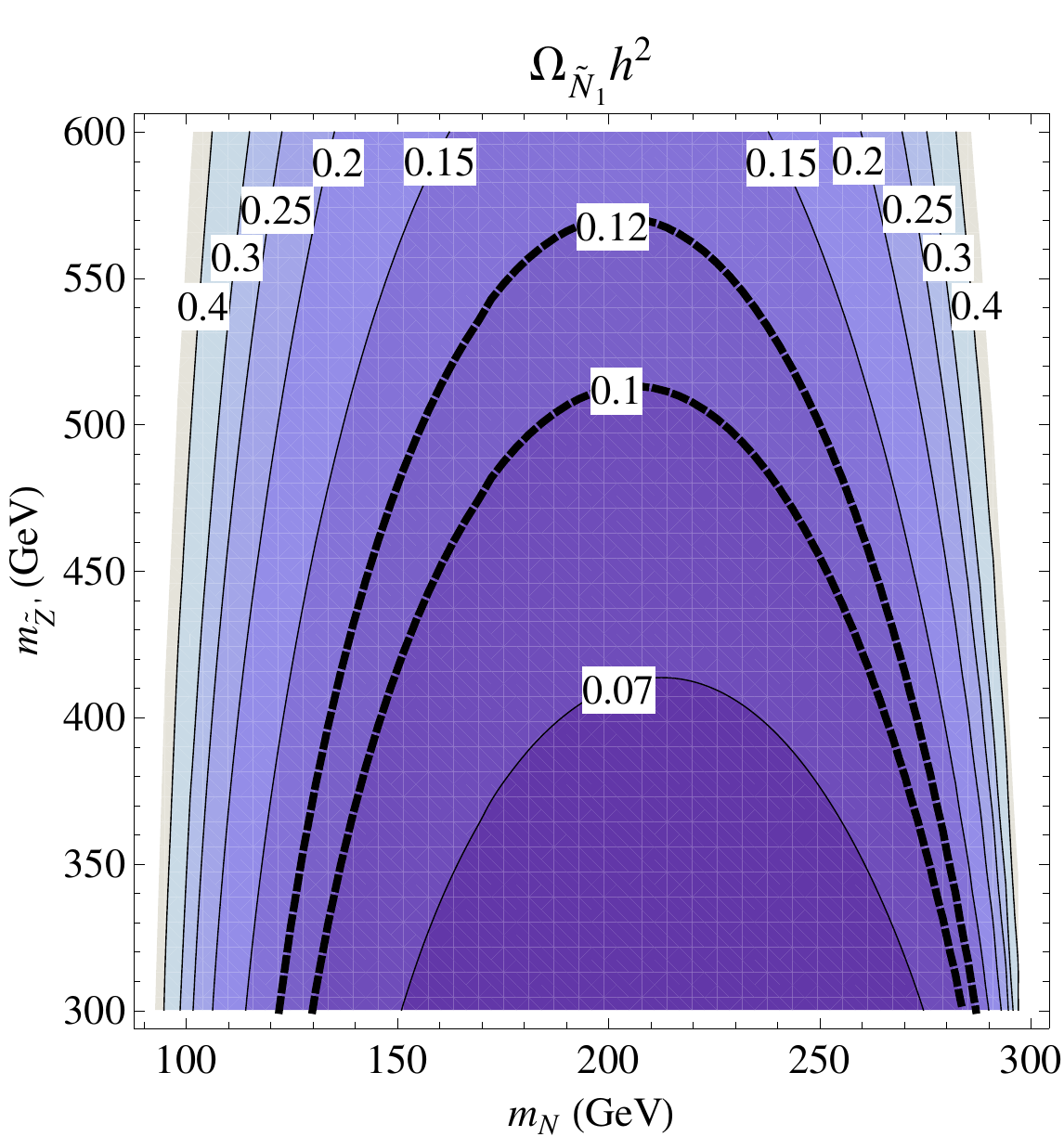}
\end{center}
\caption{Contour plots for the relic abundance of right-handed
sneutrino dark matter $\widetilde{N}_1$ in the $m_N-
m_{\widetilde{Z}'}$ plane. Left and right panels are
respectively corresponding to $\widetilde{m}_\nu = 10^{-1} $ and $10^{-5}$ eV.
The parameters $m_{\widetilde{N}_1}$ and $m_{Z'}$ are fixed
as $m_{\widetilde{N}_1} = 300$ GeV and $M_{Z'} = 1200$ GeV.
In each plane, the region between two thick dashed lines is
allowed by the recent DM relic density result.
}\label{fig:mnu}
\end{figure}
%%%%%%%%%%%%%%%%%%%%%%%%%%%%%%%%%%%%%%%%%%%%%%%%%%%%%%%%%%%%%%%%
%
%
In Figure~\ref{fig:mnu}, the thermal relic density
$\Omega_{\widetilde{N}_1}h^2$ is shown for the two representative
cases $\widetilde{m}_\nu = 10^{-1}$ and $10^{-5}$ eV in the
$m_N$--$m_{\widetilde{Z}'}$ plane to be compared with Figure~\ref{fig:mdm}.
We use the
numerical values $m_{\widetilde{N}_1} =$ 300 GeV and $M_{Z'} =
1200$ GeV for this analysis.

For smaller effective neutrino
mass, the decay rate of the right-handed neutrino is weaker, and
consequently the right-handed sneutrino DM particle is decoupled
earlier from thermal equilibrium. As a result, the relic abundance
$\Omega_{\widetilde{N}_1}h^2$ increases as
$\widetilde{m}_\nu$ decreases. This implies that smaller $\tilde Z'$ mass is
required for smaller $\tilde m_\nu$.
 One can see the tendency from the
top-middle panel of Figure~\ref{fig:mdm} and Figure~\ref{fig:mnu}.

\subsection{Dependence on $M_{Z'}$}

The
right-handed sneutrino DM $\widetilde{N}_1$ is kept in thermal equilibrium
through the interactions of the right-handed neutrino $N$  which also
depends on the $Z'$ mass as
the right-handed neutrino annihilates
to the SM fermions trough the $s$-channel $Z'$ mediation.
%
%
%%%%%%%%%%%%%%%%%%%%%%%%%%%%%%%%%%%%%%%%%%%%%%%%%%%%%%%%%%%%%%%
\begin{figure}
\begin{center}
\includegraphics[width=0.49\linewidth]{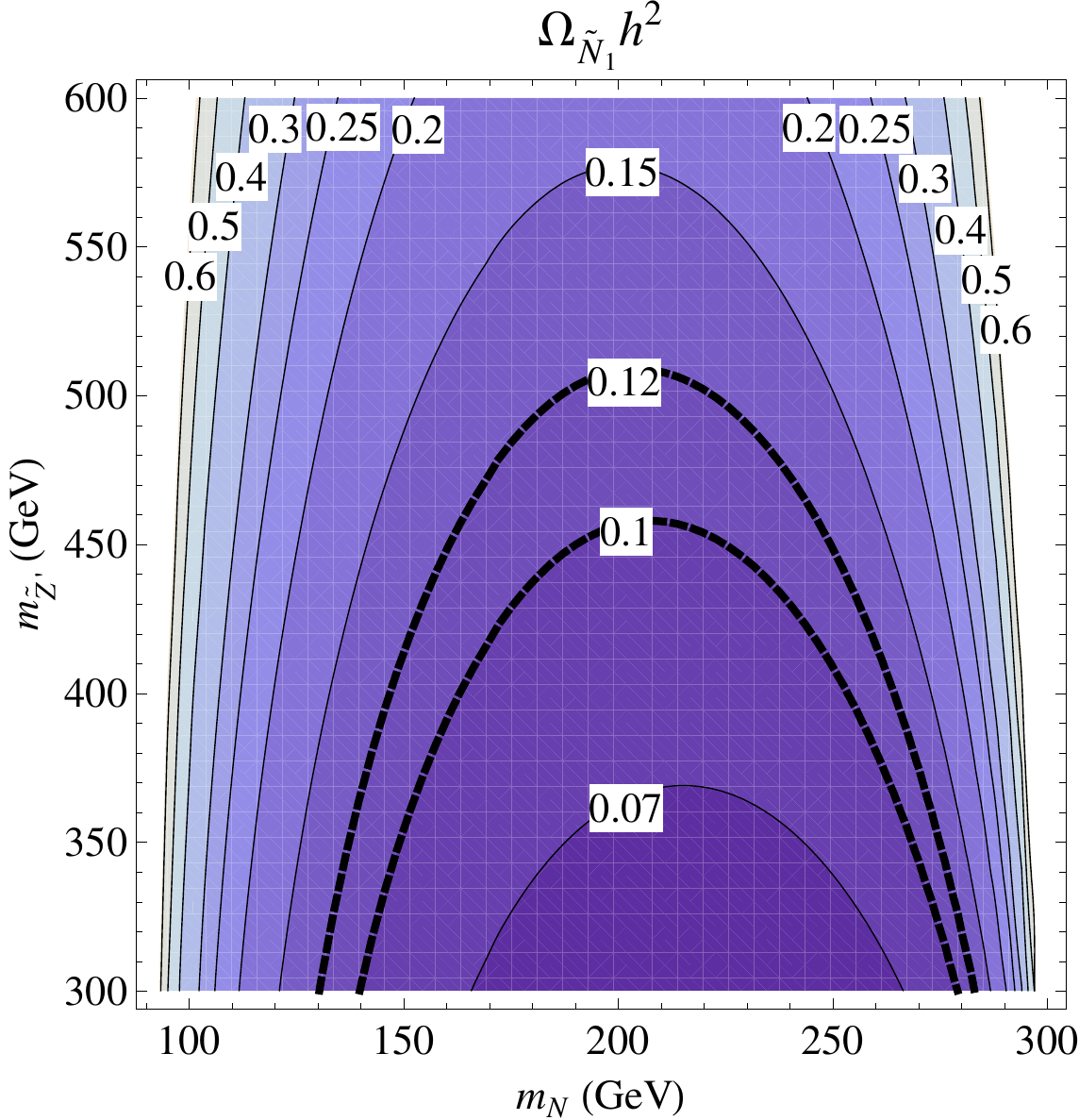}
\includegraphics[width=0.49\linewidth]{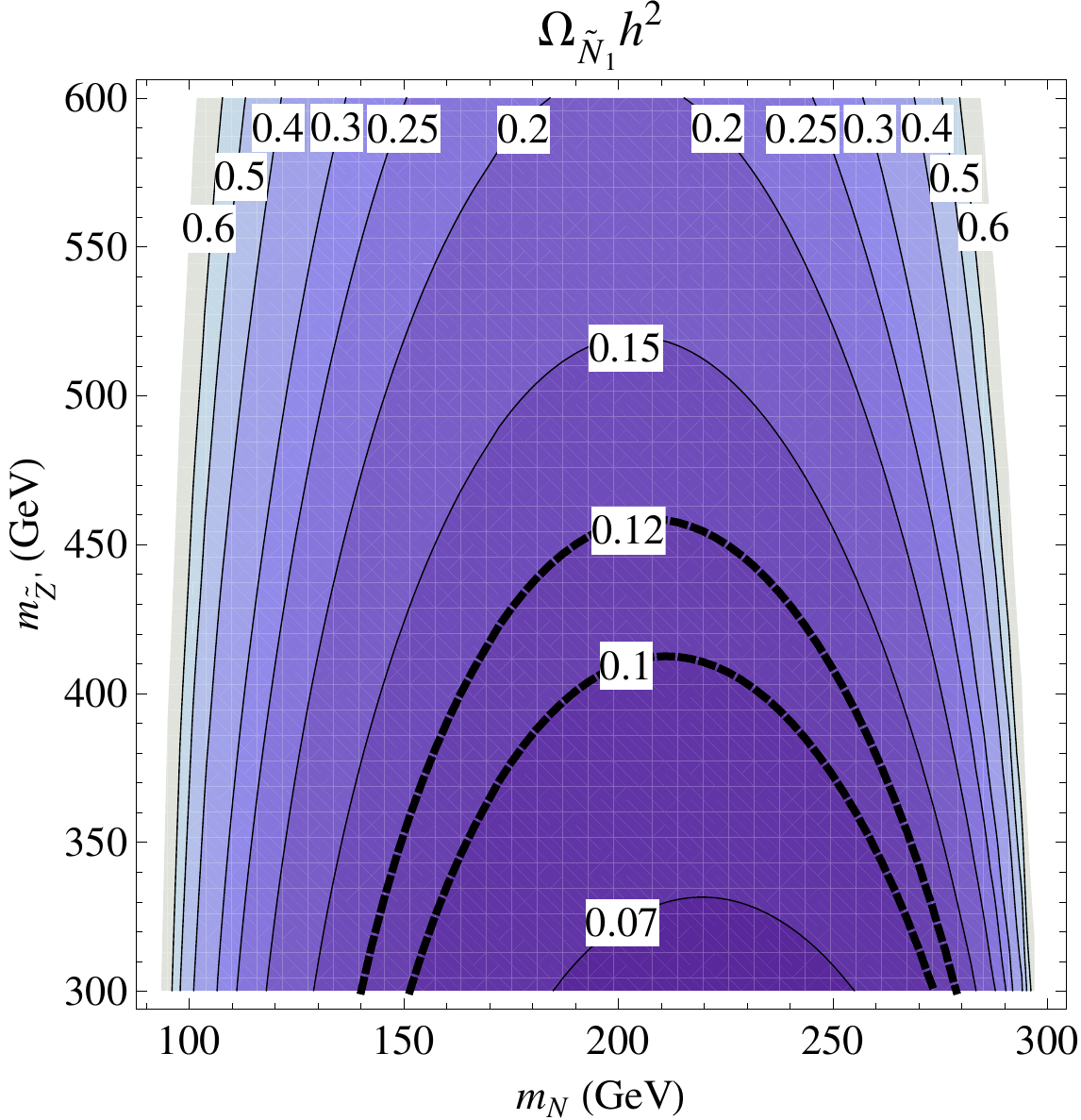}
\end{center}
\caption{Contour plots for the relic abundance of the right-handed
sneutrino dark matter $\widetilde{N}_1$ in the $m_N-
m_{\widetilde{Z}'}$ plane for the cases $M_{Z'} = $
2000 and 4000 GeV. The parameters
$m_{\widetilde{N}_1}$ and $\widetilde{m}_\nu$ are fixed as $m_{\widetilde{N}_1}
= 300$ GeV and $\widetilde{m}_\nu = 10^{-3}$ eV. The regions between two thick
dashed lines are preferred by the recent result on the DM
relic density. }\label{fig:mZchi}
\end{figure}
%%%%%%%%%%%%%%%%%%%%%%%%%%%%%%%%%%%%%%%%%%%%%%%%%%%%%%%%%%%%%%%%
%
%
In Figure~\ref{fig:mZchi}, we therefore present the thermal relic
abundance $\Omega_{\widetilde{N}_1}h^2$ in the $m_N$--$m_{\widetilde{Z}'}$ parameter space
to check the dependence of the
relic abundance $\Omega_{\widetilde{N}_1}h^2$ on the $Z'$ mass
$M_{Z'}$. In the figure, the left-panel corresponds to
$M_{Z'} =$ 2000 GeV and the right-panel to $M_{Z'} =$ 4000 GeV. In
this analysis, we use the numerical values $m_{\widetilde{N}_1} =$
300 GeV and $\widetilde{m}_\nu = 10^{-3}$ eV. The parameter space
between two thick dashed lines is also the region which is allowed
by the recent WMAP observational result on the DM relic
density~\cite{WMAP7}.

The decoupling of the right-handed sneutrino DM particle
$\widetilde{N}_1$ from thermal equilibrium is determined by the
$\widetilde{N}_1\widetilde{N}_1$ annihilation to right-handed
neutrinos through the $t$-channel $\widetilde{Z}'$ exchange if the
right-handed neutrino $N$ remains in thermal equilibrium.
Therefore, the relic density of the right-handed sneutrino DM
barely depends on the $U(1)'$ gauge boson mass $M_{Z'}$ if the
right-handed neutrino remains in thermal equilibrium by some other
interactions when the DM particle $\widetilde{N}_1$ is decoupled.
This is the case with large values of the effective neutrino mass,
$\widetilde{m}_\nu \gtrsim 10^{-3}$ eV, for which the decays and
inverse decays of the right-handed neutrino become dominant
interactions before the annihilation of $N$ freezes out. Thus, the
right-handed neutrino remains continuously in thermal equilibrium
during the freeze-out of the DM $\tilde N_1$ as can be seen
clearly from the top-left panel of Figure~\ref{fig:xf} with
$\tilde m_\nu=10^{-2}$ eV. Consequently, one can expect that the
relic abundance of $\widetilde{N}_1$ has almost no dependence on
the $Z'$ gauge boson mass. However, for smaller values of the
effective neutrino mass, $\widetilde{m}_\nu \lesssim 10^{-3}$ eV,
the annihilation of $N$ may freeze out before the decay modes of
$N$ become important, and hence the decoupling effect of the $NN$
annihilation is non-negligible. As $M_{Z'}$ increases, the $NN$
annihilation cross section becomes smaller, and thus the $NN$
annihilation decouples earlier. Because of this, the relic density
of the right-handed sneutrino DM becomes larger. Comparing the
top-middle panel of Figure~\ref{fig:mdm} and the two panels of
Figure~\ref{fig:mZchi} with $M_{Z'}=1.2, 2$ and 4 TeV, respectively,
one can find the fact that the thermal relic abundance
$\Omega_{\widetilde{N}_1}h^2$ increases as the $Z'$ gauge boson
mass increases, and thus the right dark matter density is obtained
for smaller $\tilde Z'$ mass.

\subsection{Dependence on $g'$}\label{g'dependence}

In all analysis of this paper, we use $g'=\sqrt{5/3} g_2 \tan
\theta_W \approx 0.46$ as the reference value for $U(1)'$ gauge
coupling~\cite{Langacker08}. However, in this section, we vary the
$U(1)'$ gauge coupling to show the dependence of the relic
abundance of $\widetilde{N}_1$ on this gauge coupling.
%
%
%%%%%%%%%%%%%%%%%%%%%%%%%%%%%%%%%%%%%%%%%%%%%%%%%%%%%%%%%%%%%%%
\begin{figure}
\begin{center}
\includegraphics[width=0.328\linewidth]{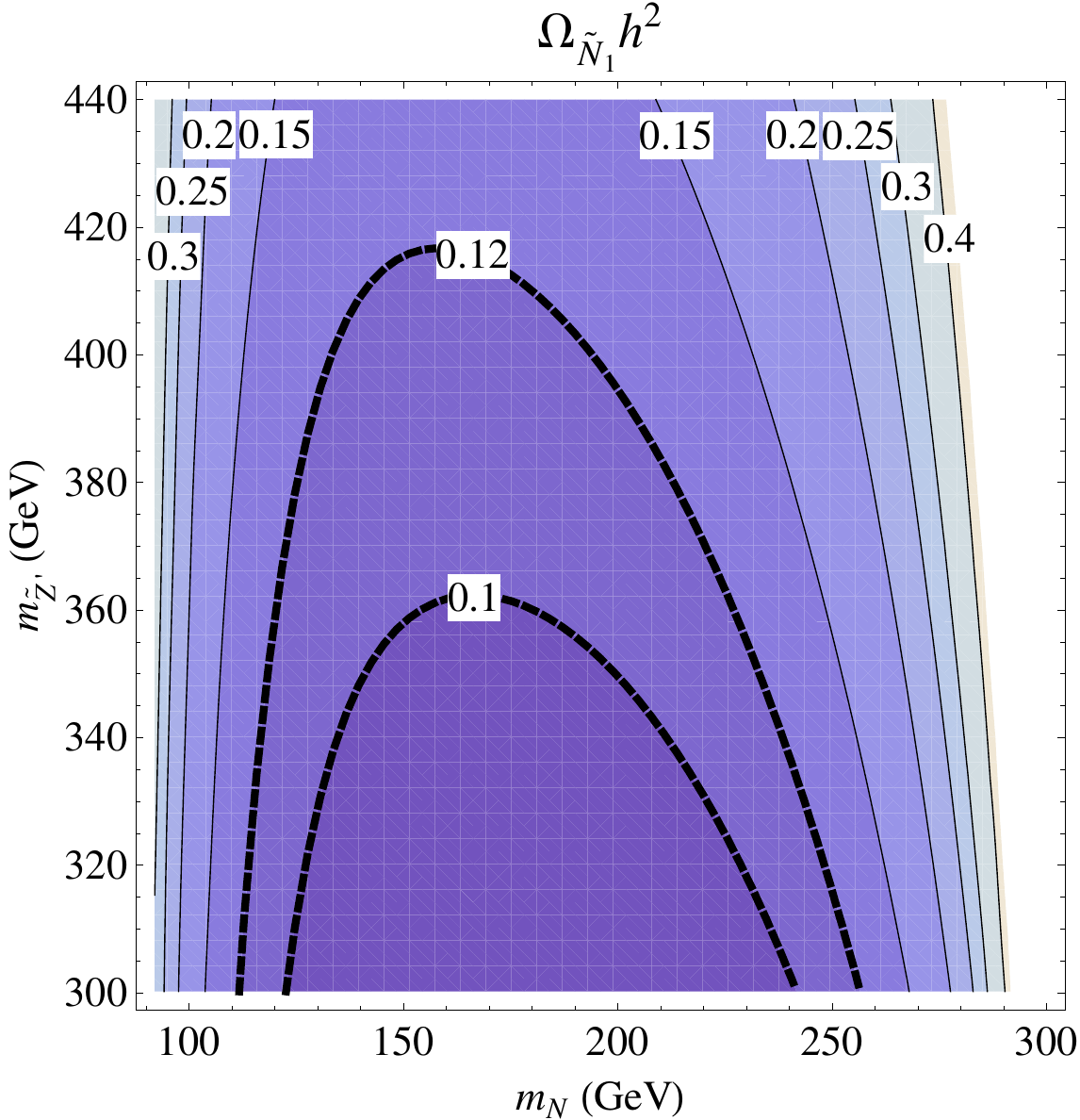}
\includegraphics[width=0.328\linewidth]{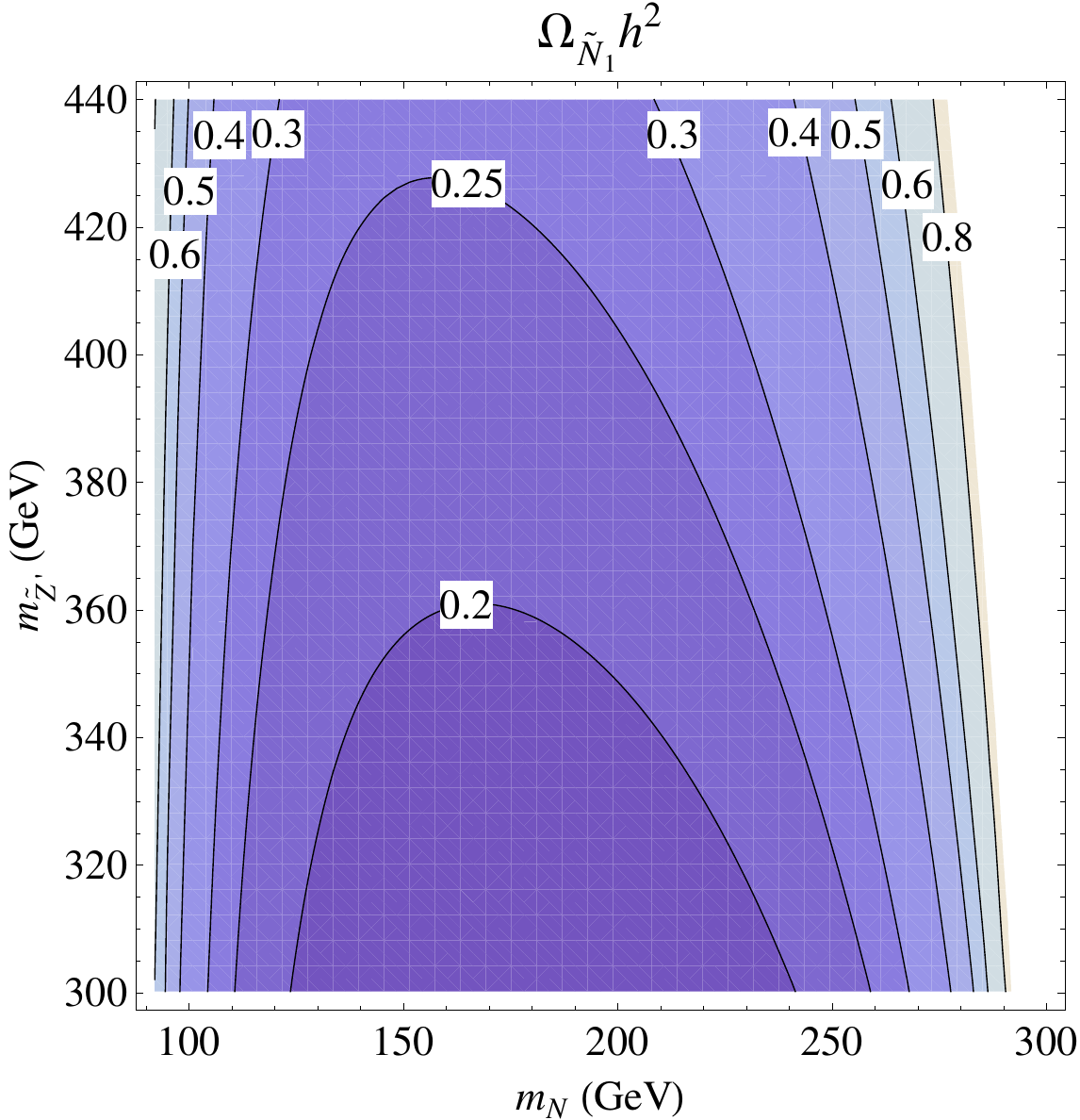}
\includegraphics[width=0.328\linewidth]{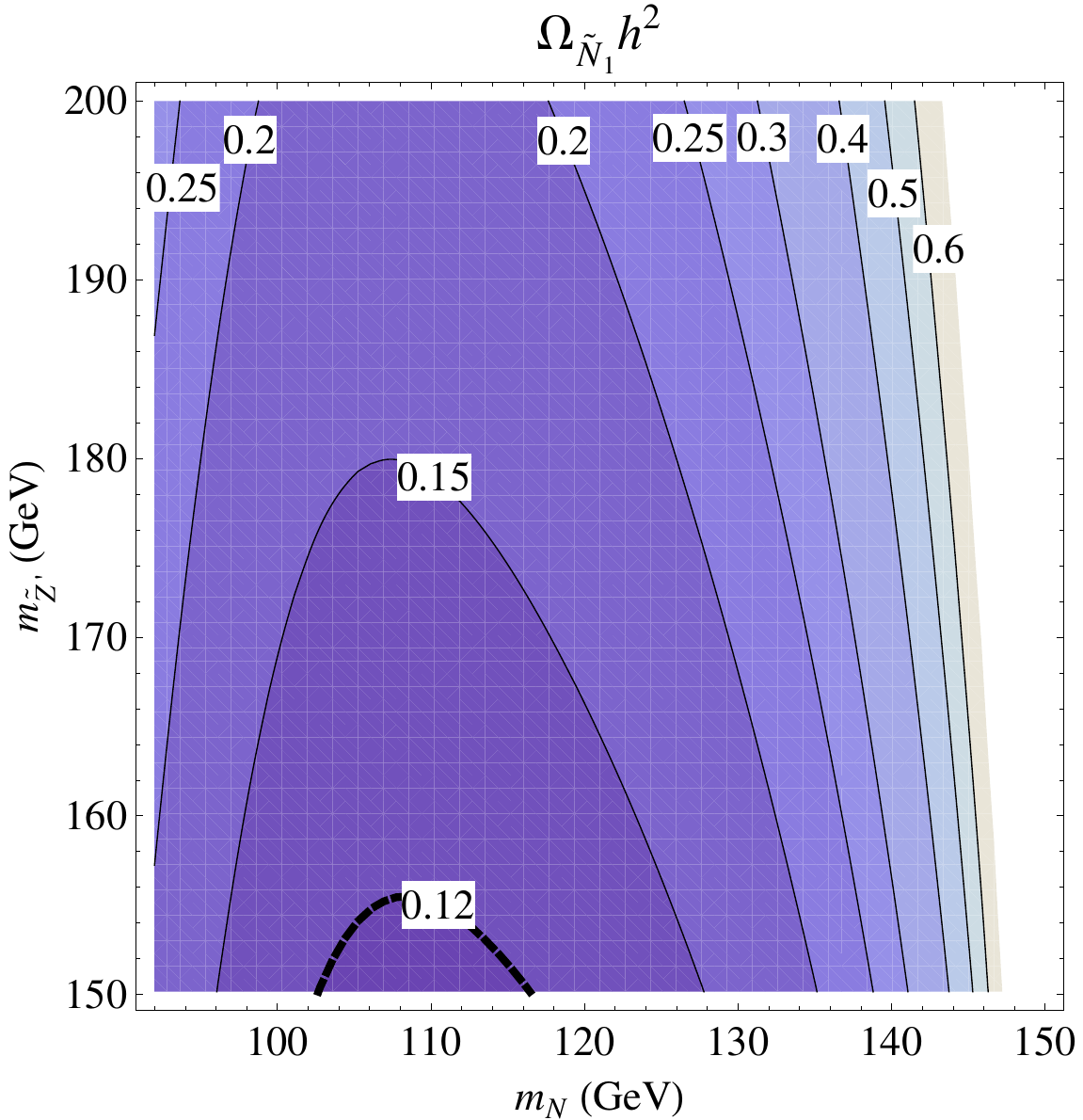}
\end{center}
\caption{Contour plots for the relic abundance of the right-handed
sneutrino dark matter $\widetilde{N}_1$ in the
$m_N-m_{\widetilde{Z}'}$ plane. The left two panels respectively
show $g'=0.3$ and $g'=0.25$ with $m_{\widetilde{N}_1} = 300$ GeV,
and the right panel shows $g'=0.2$ with $m_{\widetilde{N}_1} = 150$ GeV.
We fix the other parameters as follows: $M_{Z'} = 1200$ GeV and
$\widetilde{m}_\nu = 10^{-3}$ eV. The regions
between two thick dashed lines are allowed by the recent
DM relic density observation. }\label{fig:g'}
\end{figure}
%%%%%%%%%%%%%%%%%%%%%%%%%%%%%%%%%%%%%%%%%%%%%%%%%%%%%%%%%%%%%%%%
%
%
In the left two panels of Figure~\ref{fig:g'}, we present the
thermal relic density of $\widetilde{N}_1$ for two cases $g'=0.3$
and $g'=0.25$ with $m_{\widetilde{N}_1} = 300$ GeV; in the right
panel, for the case $g'=0.2$ with $m_{\widetilde{N}_1} = 150$ GeV.
In all the cases, we use the same numerical values
for the other parameters: $M_{Z'} = 1200$ GeV and
$\widetilde{m}_\nu = 10^{-3}$ eV. As in the previous cases, the
regions between two thick dashed lines are consistent with the
recent DM relic density observation~\cite{WMAP7}.

Obviously, for smaller $g'$, the interactions of $\widetilde{N}_1$
and $N$ become weaker, and thus the right-handed sneutrino DM is
decoupled earlier from thermal bath. Consequently, the thermal
relic density of DM increases, which can be seen from the
top-middle (top-left) panel of Figure~\ref{fig:mdm} and the left
two panels (right panel) of Figure~\ref{fig:g'}. As can be seen
from the left two panels of Figure~\ref{fig:g'}, the right-handed
sneutrino is overproduced for $g' \lesssim 0.25$ if
$\widetilde{N}_1$ is heavier than 300 GeV. When $\widetilde{N}_1$
is light enough, $m_{\widetilde{N}_1} \approx 150$ GeV, the
allowed parameter space marginally exists even for $g' = 0.2$.
Such a low limit is almost independent of the effective neutrino
mass if $\tilde m_\nu \gtrsim 10^{-3}$ eV.

As a result,  an interesting lower bound on the $U(1)'$ gauge
coupling $g' \gtrsim 0.2$ can be put by the current dark matter
relic density limit~\cite{WMAP7} in the $U(1)_\chi$ model with the
right-handed sneutrino dark matter.

\section{LHC signatures}

For the smoking gun signal of this model, we look for the
productions and decays of the new particles in the model. These include the
extra gauge boson $Z^\prime$, its superpartner $\tilde{Z}^\prime$,
the right-handed neutrino $N$ and the LSP dark
matter $\tilde{N}_{1}$ being a superpartner of $N$. The standard
search for an extra gauge boson $Z^\prime$ is through the
observation of high mass dilepton resonances from $Z' \to l^+
l^-$ \cite{lhcZp}. When the extra $U(1)'$ is associated with the
seesaw mechanism as in this model, it is also
important to look for the mode $Z' \to NN$ \cite{Keung83,Langacker84,lhcN}
to test the Majorana nature of neutrinos and the seesaw mechanism.
In this section, we will discuss the associated phenomenology of
$\tilde Z'$, which is required to be lighter than
$Z'$ as was discussed in the previous section, in parallel
with the $Z'$ phenomenology.

\subsection{Production and decay of $Z^\prime$}

%%%%%%%%%%%% Zprime*Branching Fraction in SM U(1) extn   %%%%%%%
\begin{figure}[hbt]
\begin{center}
%\vspace*{-4.2cm}
%
\includegraphics[width=0.495\linewidth]{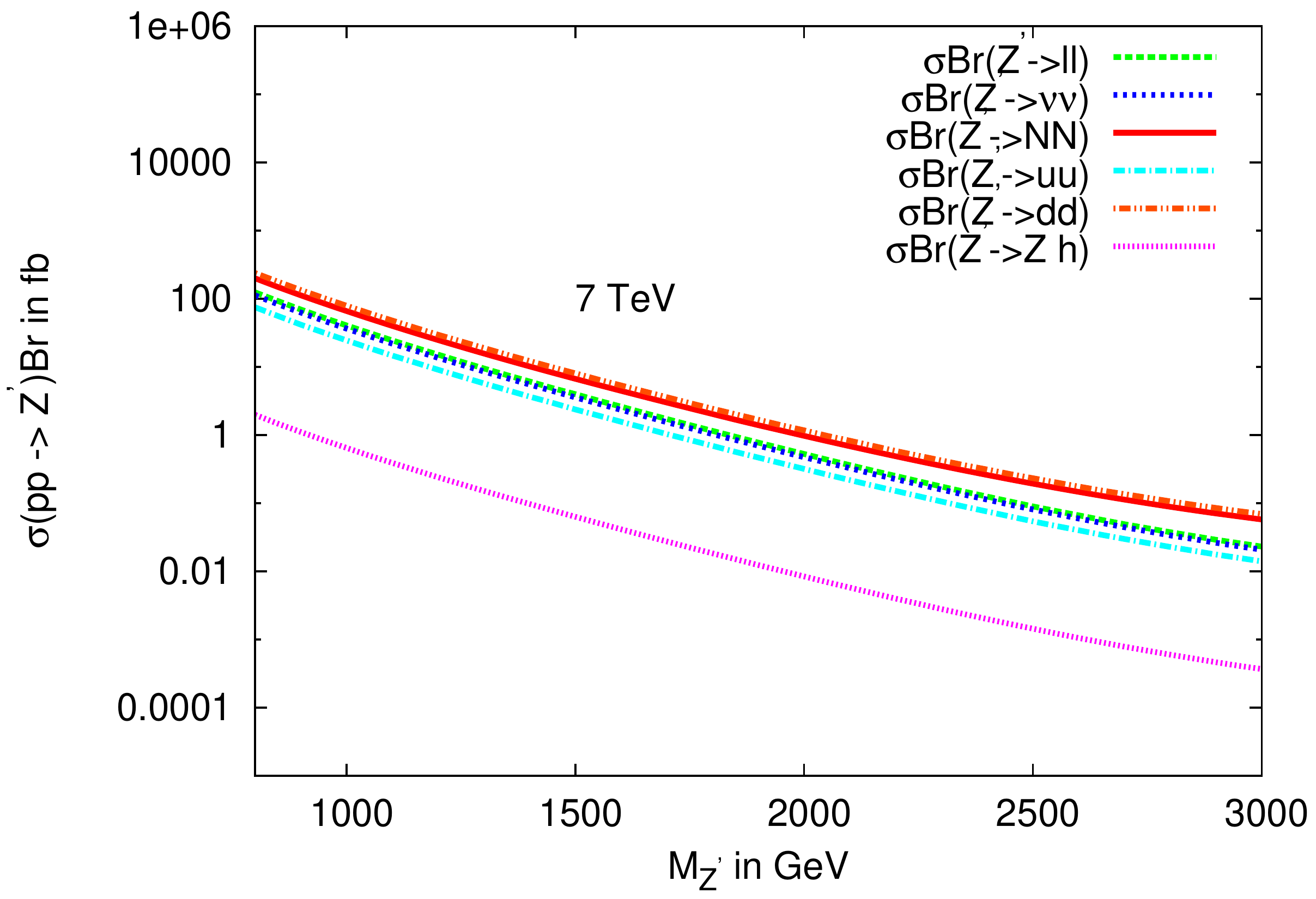}
\includegraphics[width=0.495\linewidth]{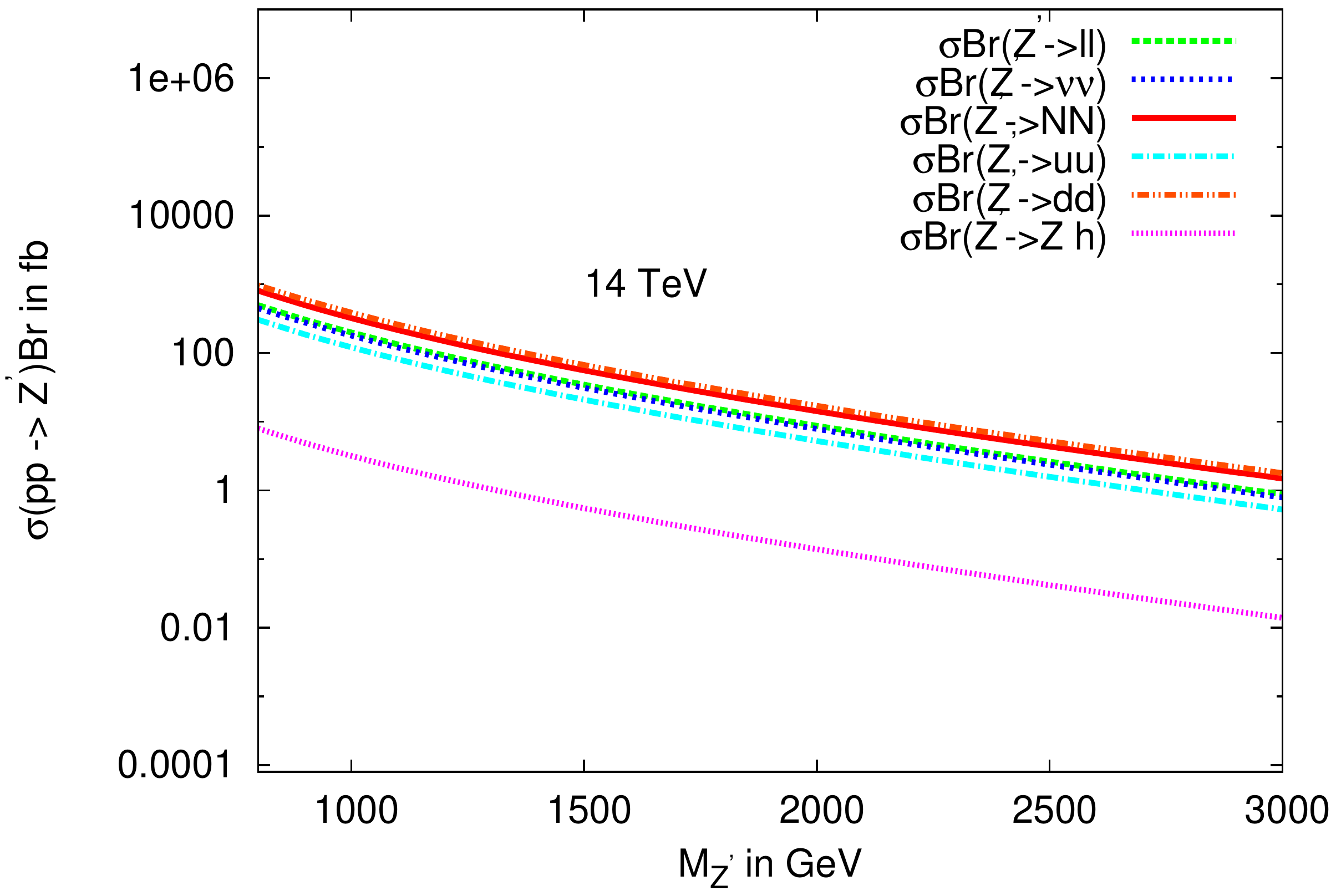}
%\hskip -15pt
%{\epsfig{file=plots/crz1sm7tev.pdf,width=6.0 cm,height=7.0cm}}
%%\vspace*{-40cm}
%\hskip -12pt
%{\epsfig{file=plots/crz1sm14tev.pdf,width=6.0 cm,height=7.0cm}}
%
%\vspace*{-16cm}
%
%\hskip -15pt
%{\epsfig{file=plots/crz1sm7tevdc.ps,width=6.0 cm,height=7.0cm,angle=-90}}
%%\vspace*{-40cm}
%\hskip -12pt
%{\epsfig{file=plots/crz1sm14tevdc.ps,width=6.0cm,height=7.0cm,angle=-90}}
%
%\vspace*{-16cm}
\caption{The $Z'$ production cross-section times the branching
ratios for the SM channels including right-handed
neutrinos. Two panels are for $\sqrt{\hat{s}}=$ 7 and 14 TeV, respectively,
with $\tan\beta=10$.}
\label{zpprod}
\end{center}
%\vspace*{-0.5cm}
\end{figure}
%%%%%%%%%%%%%%%%%%%%%%%%%%%%%%%%%%%%%%%%%%%%%%%%%%%%%%%%%

Single $Z^\prime$ produced at the LHC like the Standard Model
$Z$ can be observed through its decay modes.  All the possible
decay channels are shown in Table~\ref{couplzp} for the supersymmetric
$U(1)_\chi$ model. Our main focus is the production of the right-handed
neutrinos through the process $pp \to Z' \to NN$ which can be compared
with the standard $Z'$ discovery channel $pp \to Z' \to l^\pm l^\mp$.
In this paper, we assume that the $Z'$ decays to sfermions are not
allowed kinematically. In Figure~\ref{zpprod}, we show the
production cross-section for each SM decay channel of $Z'$ including
right-handed neutrinos in the decoupling limit of non-SM Higgs bosons.
The mass of the right-handed neutrino was taken to be 300 GeV for the
set of plots. The renormalization/factorization scale was chosen to be
$\sqrt{\hat{s}}$ and CTEQ6L \cite{Pdf} was taken as PDF for the
cross-section calculation. Apart from the dilepton and $NN$ channels,
the Higgs modes could also be interesting. For the quark, lepton and
right-handed neutrino modes in Figure \ref{zpprod}, we have shown only
one flavor contributions. One can see that the down type quark mode
has larger cross-section than the up type quark as well as the leptonic
modes, which is consistent with the charges given in Eq.~(\ref{chargeqn}).
By the same reason, the $NN$ production cross-section is the largest,
which enhances the Majorana neutrino signatures (see Subsection 4.3)
compared to other $U(1)'$ models. Concerning the Higgs contribution,
we have considered only the SM Higgs which leads to the $Zh$ mode. This
has a very low cross-section for large $\tan\beta$ as the coupling is
proportional to $\cos(\alpha+\beta)$ which is $\sin(2\beta)$ in the
decoupling limit. Thus, for low $\tan{\beta}$, the Higgs cross-section
gets enhanced and comparable to the fermionic modes.
%
% Also the ratio of the production rates of $pp\to Zh/H$
%to $pp\to \nu\nu$ is roughly, $\sim \frac{\cos^2(\alpha+\beta)/\sin^2(\alpha+\beta)}{9}$.
%Thus going from $\tan{\beta}=10$ to $\tan{\beta}=2$ the $pp\to Zh$
%production rates gets an enhancement by a factor $\sim 5.5$
%where as the $pp \to ZH$ gets a reduction by $\sim 0.14$.}
%
Let us also remark that the production cross-section of the standard $Z'$ discovery channel,
$Z' \to l^\pm l^\mp$,  depends on $\tan\beta$ and on the number of new channels, e.g.,
with Higgs bosons or sfermions that can be kinematically allowed in our model.
In the case of a $U(1)'$ extension of the non-supersymmetric Standard Model, the model dependence comes only from the number of right-handed neutrino channels.

%%%%%%%%%%%% Zprime*Branching Fraction in BSM U(1) extn  %%%%%%%
\begin{figure}[hbt]
\begin{center}
%\vspace*{-4.2cm}
\includegraphics[width=0.495\linewidth]{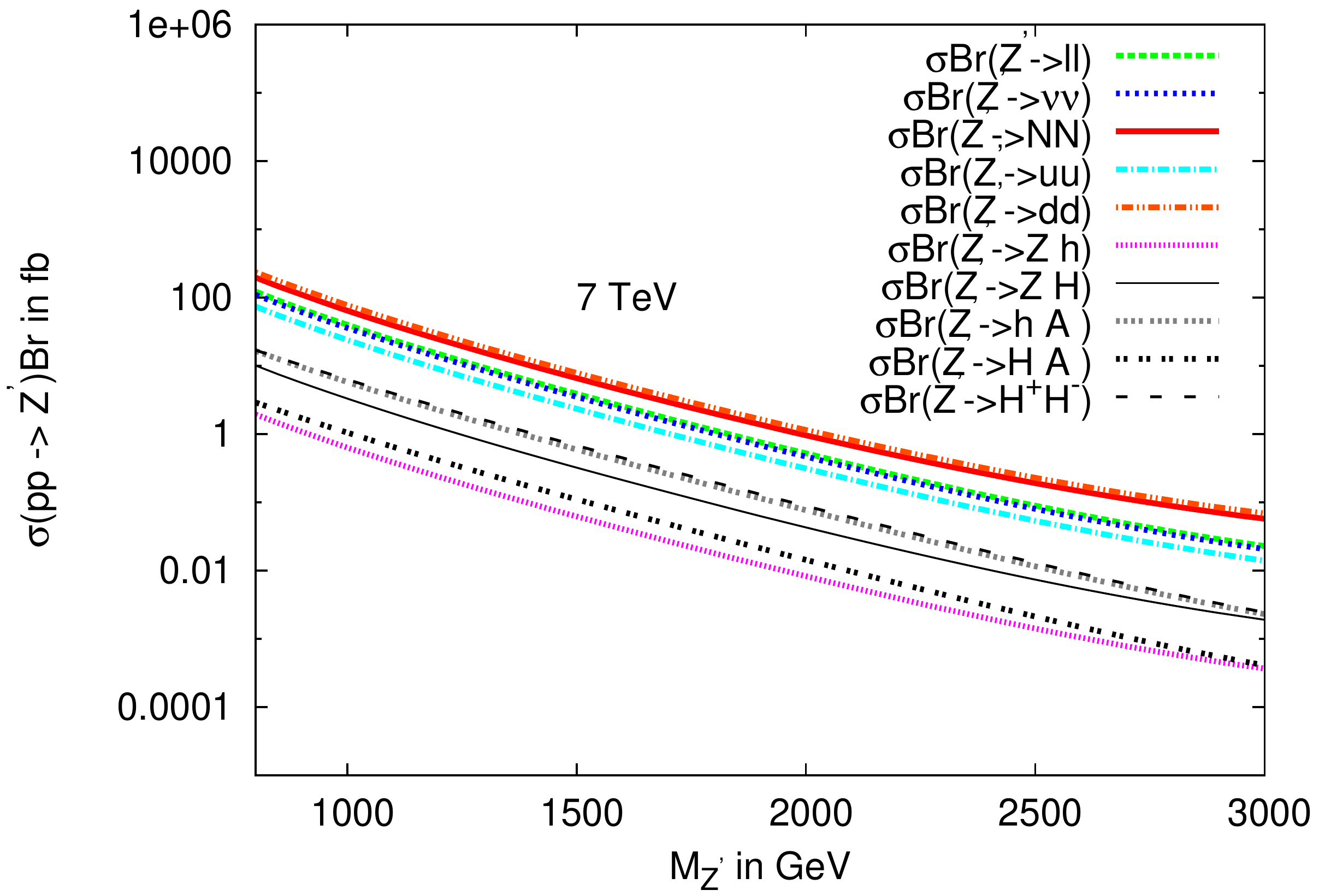}
\includegraphics[width=0.495\linewidth]{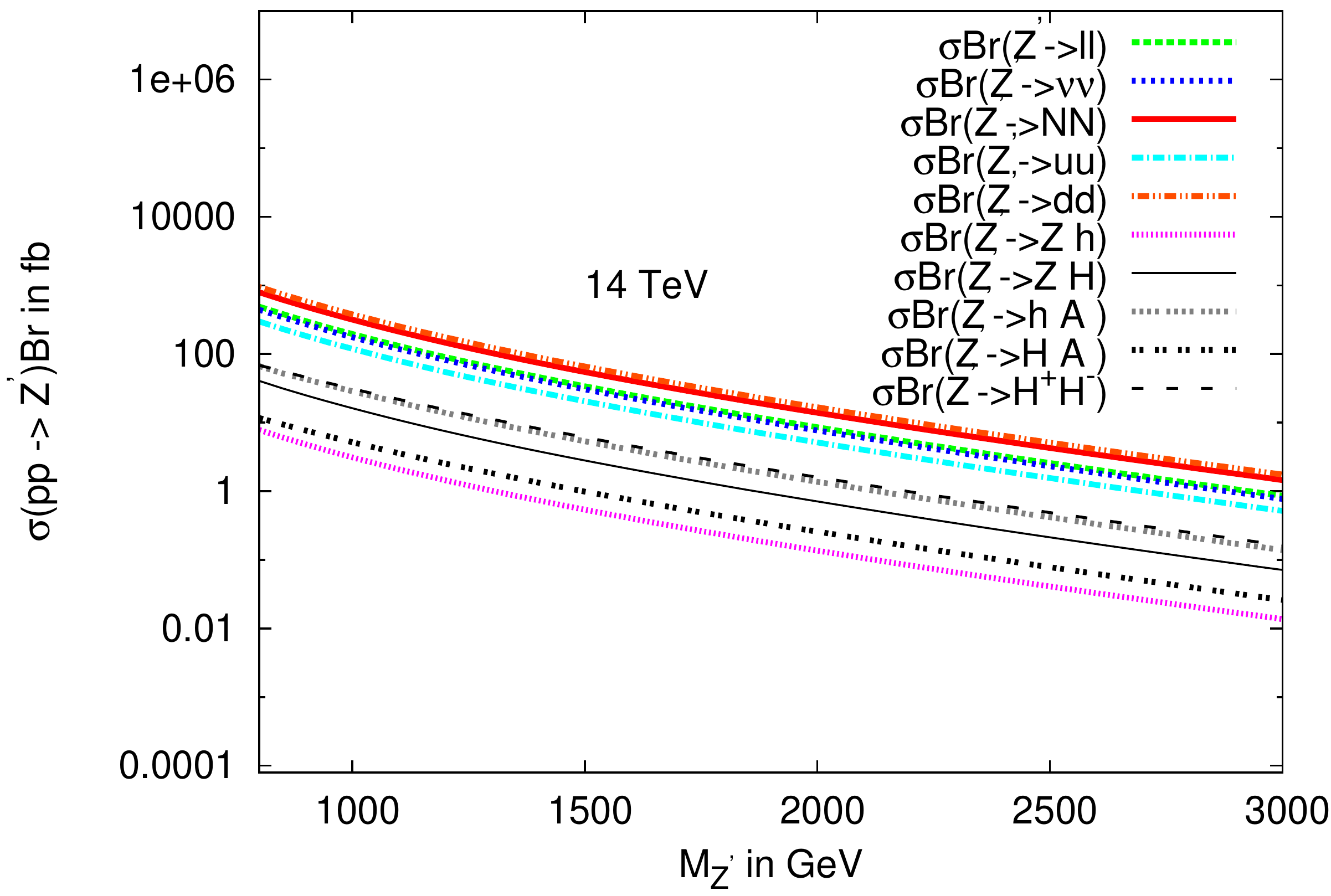}
%
%\hskip -15pt
%{\epsfig{file=plots/crz17tev.pdf,width=6.0 cm,height=7.0cm}}
%%\vspace*{-40cm}
%\hskip -12pt
%{\epsfig{file=plots/crz114tev.pdf,width=6.0cm,height=7.0cm}}
%
%\vspace*{-16cm}
\caption{Same as in Figure~\protect\ref{zpprod} but with non-decoupling heavy Higgs bosons.}
\label{zpprodss}
\end{center}
%\vspace*{-0.5cm}
\end{figure}
%%%%%%%%%%%%%%%%%%%%%%%%%%%%%%%%%%%%%%%%%%%%%%%%%%%%%%%%%

Figure \ref{zpprodss} describes the case with all the possible
Higgs decay modes, which includes $Z^\prime$ decays to $HZ$,
$hA$, $HA$ and $H^\pm H^\mp$. For this study, we have taken
the CP-odd Higgs to be non-decoupled , i.e. $m_A=162.2$ GeV,
which makes the charged Higgs mass 180 GeV and the heavy neutral
Higgs 164 GeV. As can be seen in Figure~\ref{zpprodss},
the production cross-sections are not promising at least for the
case of center of mass energy of 7 TeV. In the 14 TeV case,
$H^\pm H^\mp$ and $hA$ can approach the production cross-section
$\sim 100$ fb at $\tan{\beta}=10$. The other Higgs channels, $HA$,
$hZ$ and $HZ$, do not have enough effective-cross section at 14 TeV.
At $\tan{\beta}=10$, $\sin(\alpha+\beta)\simeq 1$, which makes
the $h$--$A$--$Z^\prime$ and $Z^\prime$--$H^\pm$--$H^\mp$ couplings
almost same (see Table \ref{couplzp}) which is also reflected in
Figure \ref{zpprodss}. The ratio of the production cross-section
between $hA$ and $HA$ is proportional to $\left(\frac{\sin(\alpha+\beta)}{\cos(\alpha+\beta)}\right)^2$ and
the same is the case for $ZH$ and $Zh$, as is expected from the
nature of their couplings given in Table~\ref{couplzp}. It is also
to be noticed  that Higgs pair production channels  have more
production rates compared to associated gauge boson channels due to
the momentum dependent couplings of the formers. Given the suppressed
Higgs channels, the production cross-sections for the $NN$ and $ll$
modes remain almost same as in the decoupling limit.

\subsection{Production and decay of $\tilde{Z}^\prime$}

\begin{figure}[tbh]
\begin{center}
%\vspace*{-4.2cm}
%
\includegraphics[width=0.7\linewidth]{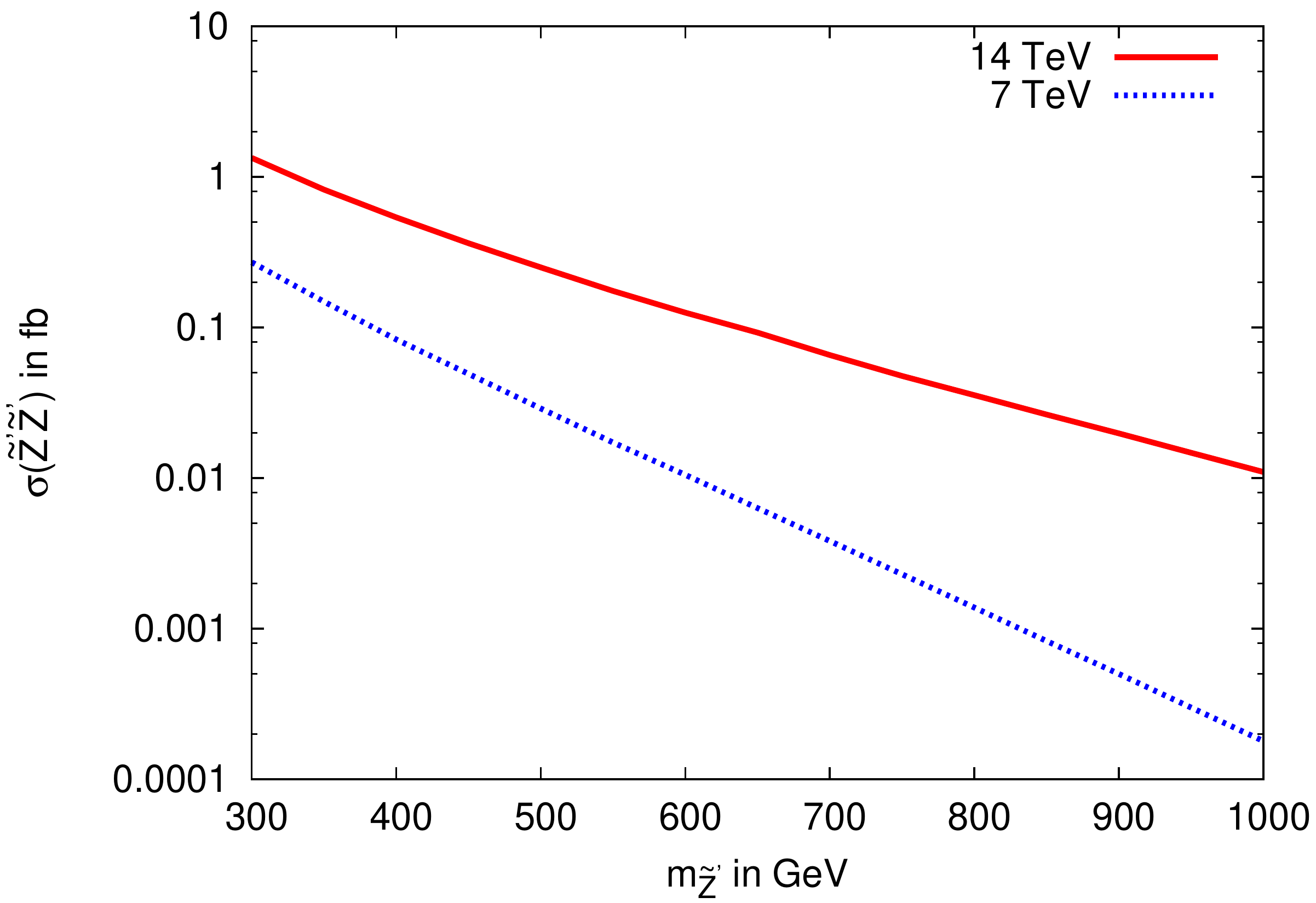}
%
%\hskip -15pt
%{\epsfig{file=plots/crosssz1.pdf,width=8.0 cm,height=10.0cm}}
%%\vspace*{-40cm}
\caption{Variation of $\tilde{Z}^{\prime}\tilde{Z}^{\prime}$ production cross-section.}\label{crossZ1}
\end{center}
%\vspace*{-0.5cm}
\end{figure}

Unlike $Z^\prime$ whose mass is strongly constrained by the precision
electroweak data, its super-partner $\tilde{Z}^\prime$ can be light,
as required in the model of the right-handed sneutrino dark matter.
Thus, it can lead to interesting phenomenology related to the seesaw
mechanism even though $Z'$ turns out to be too heavy to be produced
at the LHC.Let us first try to  look for the direct pair production
of $\tilde{Z}^\prime$ at the LHC. In Figure \ref{crossZ1}, we estimate
the pair production rate, \textit{i.e.} $pp \to \tilde{Z}^\prime\tilde{Z}^\prime$,
with the variation of $m_{\tilde{Z}^{\prime}}$ at the LHC for
the center of mass energy of 7 TeV and 14 TeV. The renormalization/factorization
scale was chosen to be $\sqrt{\hat{s}}$ and CTEQ6L \cite{Pdf} was
taken as PDF for the cross-section calculation as was in the previous case.
The rates appear to be very low for both 7 and 14 TeV cases, due
to the fact that only $t$-channel  electroweak diagrams contribute to
the process. This results in the production cross-section $\sim $ fb
even for the light $\tilde{Z}^{\prime}$ .

%%%%%%%%%%%%Production cross-section of ~Z_{B-L}~Z_{B-L}%%%%%%%
%%%%%%%%%%%%%%%%%%%%%%%%%%%%%%%%%%%%%%%%%%%%%%%%%%%%%%%%%%%%

However, let us note that the LHC being a machine with huge gluon
flux, the strongly interacting supersymmetric particles namely
squarks and gluino can be copiously produced. Cascade decays of
such supersymmetric colored particles could then be a good source of
superpartners of the electroweak particles. In particular, squarks
decay through the direct electroweak couplings to quarks and neutralinos
or charginos, and gluinos decay through the two-body decay to squarks
and quarks or through three-body decay to quark pair and charginos
or neutralinos. An interesting application of such supersymmetric
cascade decays to the Higgs production has been studied \cite{Higgscas}
in the context of MSSM.

%%%%%%%%%%% Strong Production at 14 and 7 TeV  %%%%%%%%%%%%%
\begin{figure}[bht]
\begin{center}
%\vspace*{-4.2cm}
%
\includegraphics[width=0.7\linewidth]{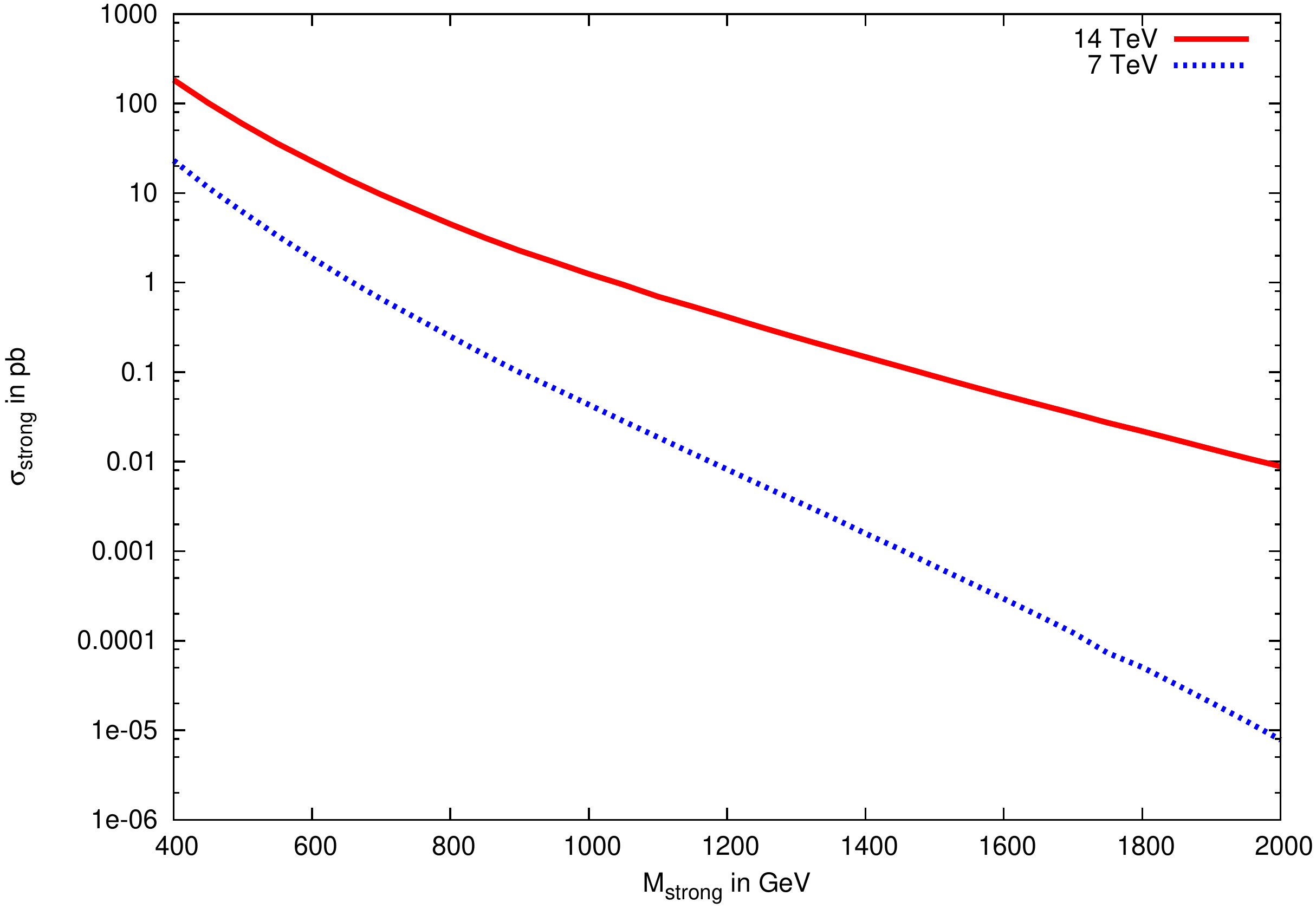}
%
%\hskip -15pt
%{\epsfig{file=plots/strongprod.pdf,width=8.0cm,height=10.0cm}}
%
%\vspace*{-16cm}
\caption{Variation of strong production cross-section.}\label{crosstrn}
\end{center}
%\vspace*{-0.5cm}
\end{figure}
%%%%%%%%%%%%%%%%%%%%%%%%%%%%%%%%%%%%%%%%%%%%%%%%%%%%%%%%%%%%%

%%%%%%%%%%% Branching Fraction  ~q-->q ~Z^\prime  %%%%%%%
\begin{figure}[hbt]
\begin{center}
%\vspace*{-4.2cm}
%
\includegraphics[width=0.495\linewidth]{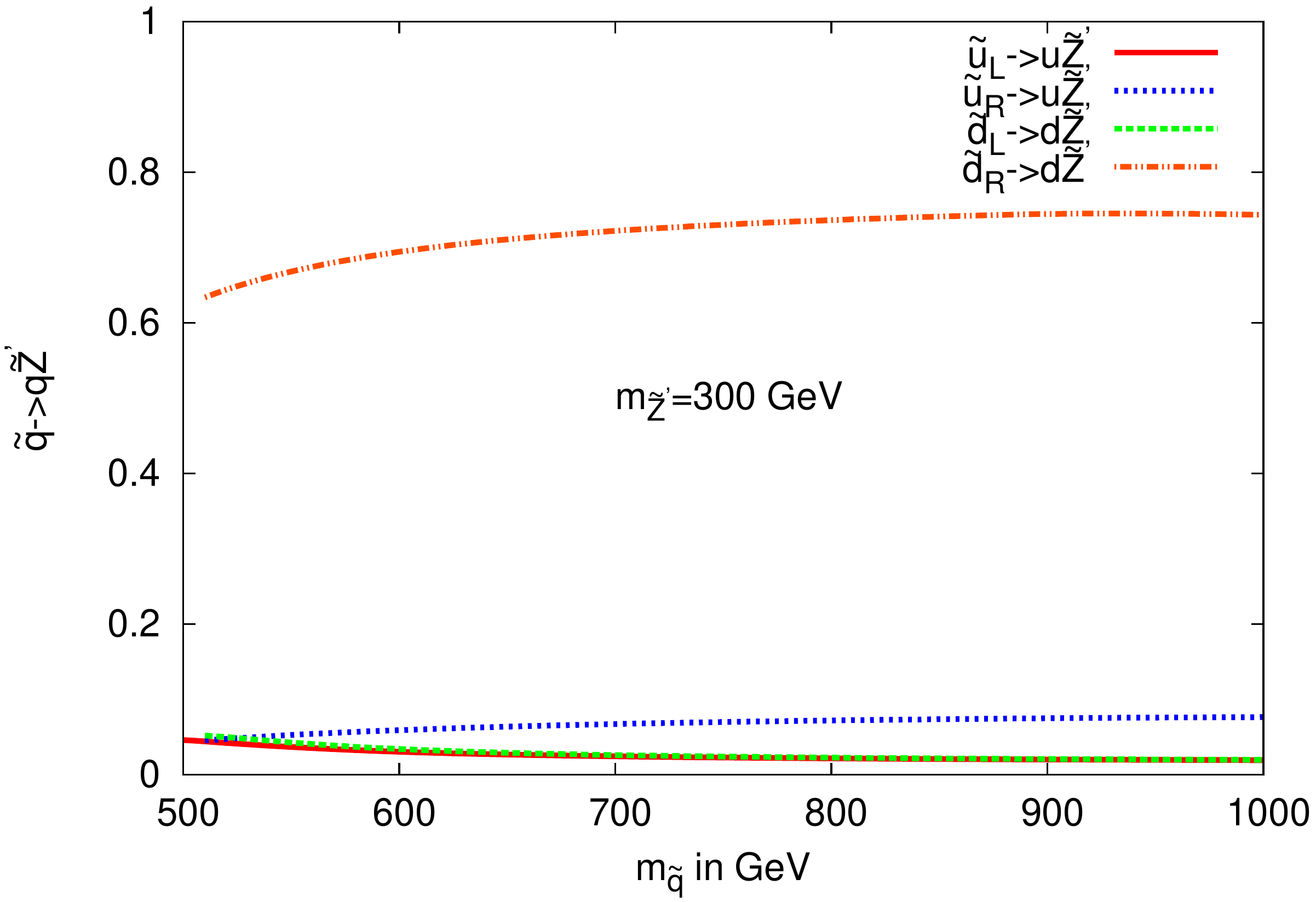}
\includegraphics[width=0.495\linewidth]{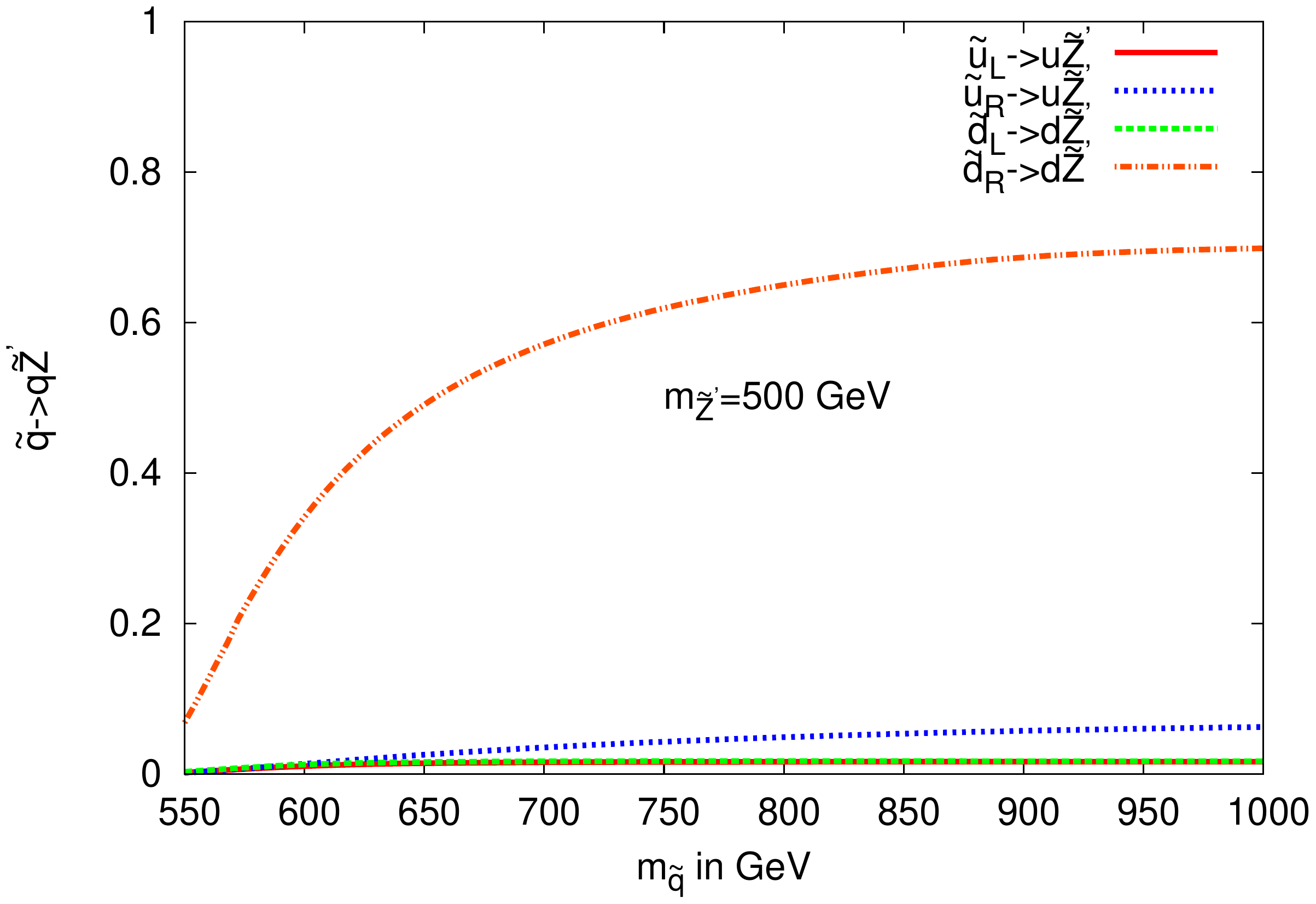}

\vspace*{0.1cm}

\includegraphics[width=0.49\linewidth]{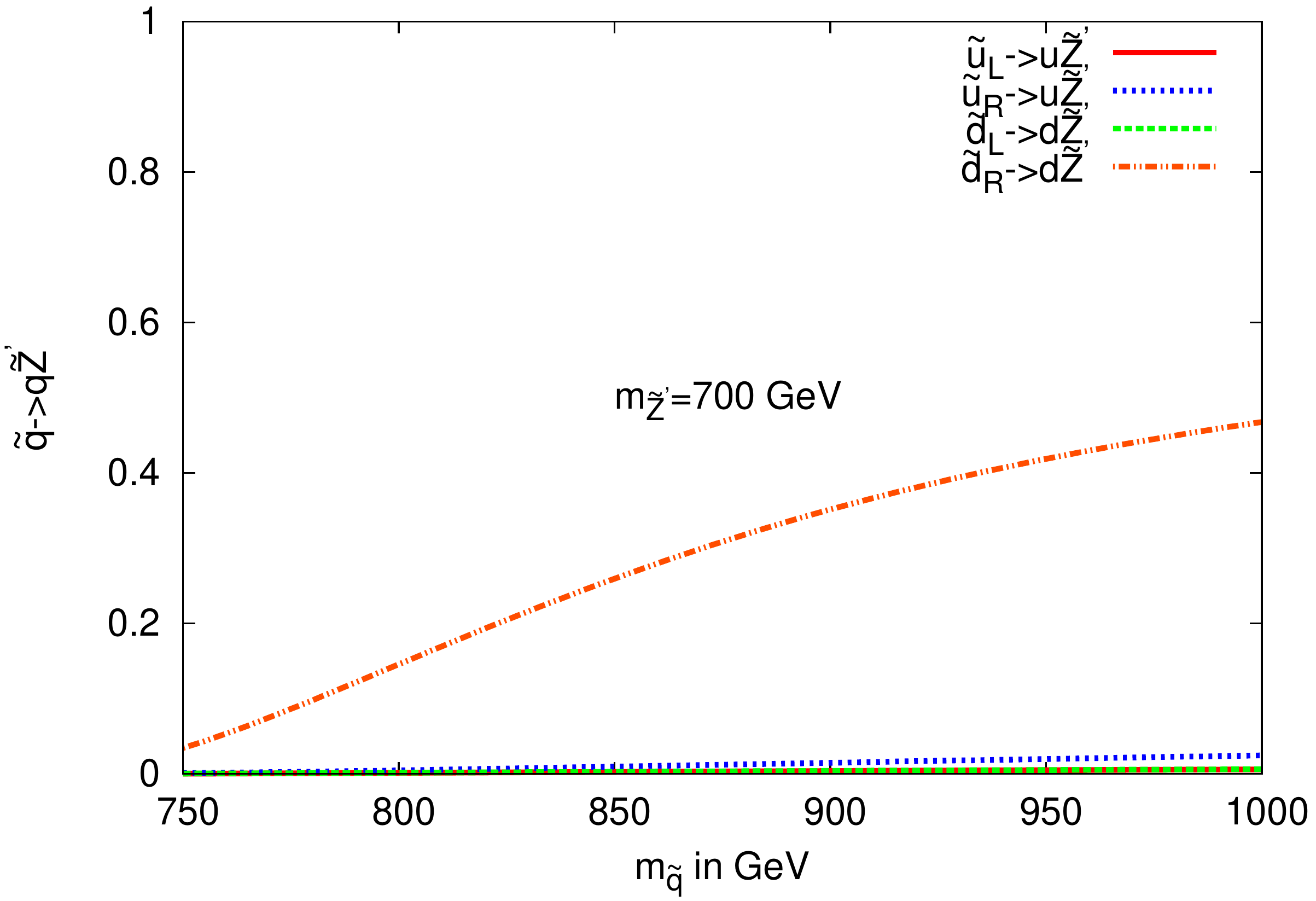}
%
%\hskip -15pt
%{\epsfig{file=plots/sqsz300.pdf,width=6.0 cm,height=7.0cm}}
%%\vspace*{-40cm}
%\hskip -12pt
%{\epsfig{file=plots/sqsz500.pdf,width=6.0cm,height=7.0cm}}
%
%\hskip -12pt
%{\epsfig{file=plots/sqsz700.pdf,width=6.0cm,height=7.0cm}}
%
%\vspace*{-16cm}

\caption{Variation of Br($\tilde{q}\to q\tilde{Z}^\prime$) with $M_{\rm strong}$ for $m_{\tilde{Z}^\prime}=$ 300, 500, and 700 GeV respectively.}\label{brsZ1}
\end{center}
%\vspace*{-0.5cm}
\end{figure}
%%%%%%%%%%%%%%%%%%%%%%%%%%%%%%%%%%%%%%%%%%%%%%%%%%%%%%%%%%%%%%

Figure \ref{crosstrn} describes the variation of strong production rate with
the common mass $M_{\rm strong}$ of squarks and gluinos for the center of
mass energy of 7 TeV and 14 TeV including the contributions from only first
two generations of squarks.
%The production cross-sections at $M_{\rm strong}=600$ GeV are 22.6 pb
%and 1.9 pb for $\sqrt{S}=14$ and 7 TeV respectively.
For $M_{\rm strong}=1$ TeV, the cross-sections are 43 fb and 1.3 pb for
the 7 and  14 TeV center of mass energy, respectively. The inclusion of
third generation will of course enhance the production rate. This can be
compared to the $Z'\to NN$ production cross-sections for $M_{Z'}=1$ TeV:
70 fb and 350 fb for the 7 and 14 TeV for one generation of right handed
neutrino, respectively. Inclusion of three generation will increase the
production cross-section by factor 3. The numbers of $\tilde Z'$ and $Z'$
events will sensitively depend on the masses of squarks/gluinos and $Z'$.
The squarks and gluinos can now decay to $\tilde{Z}^{\prime}$.
The branching fraction to $\tilde{Z}^{\prime}$ depends on the mass parameters and
on the coupling with left and right-handed
up and down type quarks as shown in Eq.~(\ref{chargeqn}).
Figure \ref{brsZ1} describes the $\tilde{q}\to q \tilde{Z}^\prime$
branching fraction  with the variation of the  squark mass for
$m_{\tilde{Z}^{\prime}}=$ 300 GeV, 500 GeV and 700 GeV.
From the figure, we can see that the $\tilde{d}_R\to d \tilde{Z}^\prime$
has larger branching fraction compared to other quark modes as is
evident from Eq.~(\ref{chargeqn}).
%that the squark decay branching fraction to $\tilde{Z}^\prime$
%depends on up-down type flavor and left-right
% handedness.
The charge corresponding to $d_R$ type quark
is $3/2\sqrt{10}$ which is the largest among quarks.
Note that the branching ratio of $\tilde q \to q \tilde Z'$ reaches about 70\%
unless the kinematic suppression is applied.  This will lead to a sizable number of $\tilde Z'$ produced from the cascade decays of squarks and gluinos.  In the case of $\tilde Z'$ being the next lightest supersymmetric particle (NLSP), all of the pair produced strong particles will end up with a pair of $\tilde Z'$, i.e. the effective branching fraction is 100 \%.

The decay of $\tilde{Z}^{\prime}$  can give rise to various
final states depending on the possible decay modes that are open: that is,
$\tilde Z' \to N \tilde N, l \tilde l, H \tilde H$ and $S_{1,2} \tilde S_{1,2}$.
Here $H$ and $\tilde H$ denote any type of Higgs bosons and Higgsinos shown in Table~\ref{couplzp}.
The $U(1)'$ Higgs bosons, $S_{1,2}$, or Higgsinos, $\tilde S_{1,2}$, are expected to be as heavy as $Z'$ and thus heavier than $\tilde Z'$. Assuming the Higgsinos heavier than $\tilde Z'$, we will concentrate on the first two decay channels in this paper. Note again that the $\tilde Z' \to N \tilde N$ mode is the dominant mode given the $U(1)_\chi$ charge assignment (\ref{chargeqn}) and becomes the unique one if $\tilde Z'$ is the NLSP and $\tilde N$ is the LSP, $\tilde N_1$. In the next subsection, we will focus on the mode $\tilde Z' \to N \tilde N_1$.

\subsection{Signatures of the seesaw and displaced Higgses}

%%%%%%%%%%%% Branching Fraction  N-->2*x  %%%%%%%
\begin{figure}[hbt]
\begin{center}
%\vspace*{-4.2cm}
%
\includegraphics[width=0.495\linewidth]{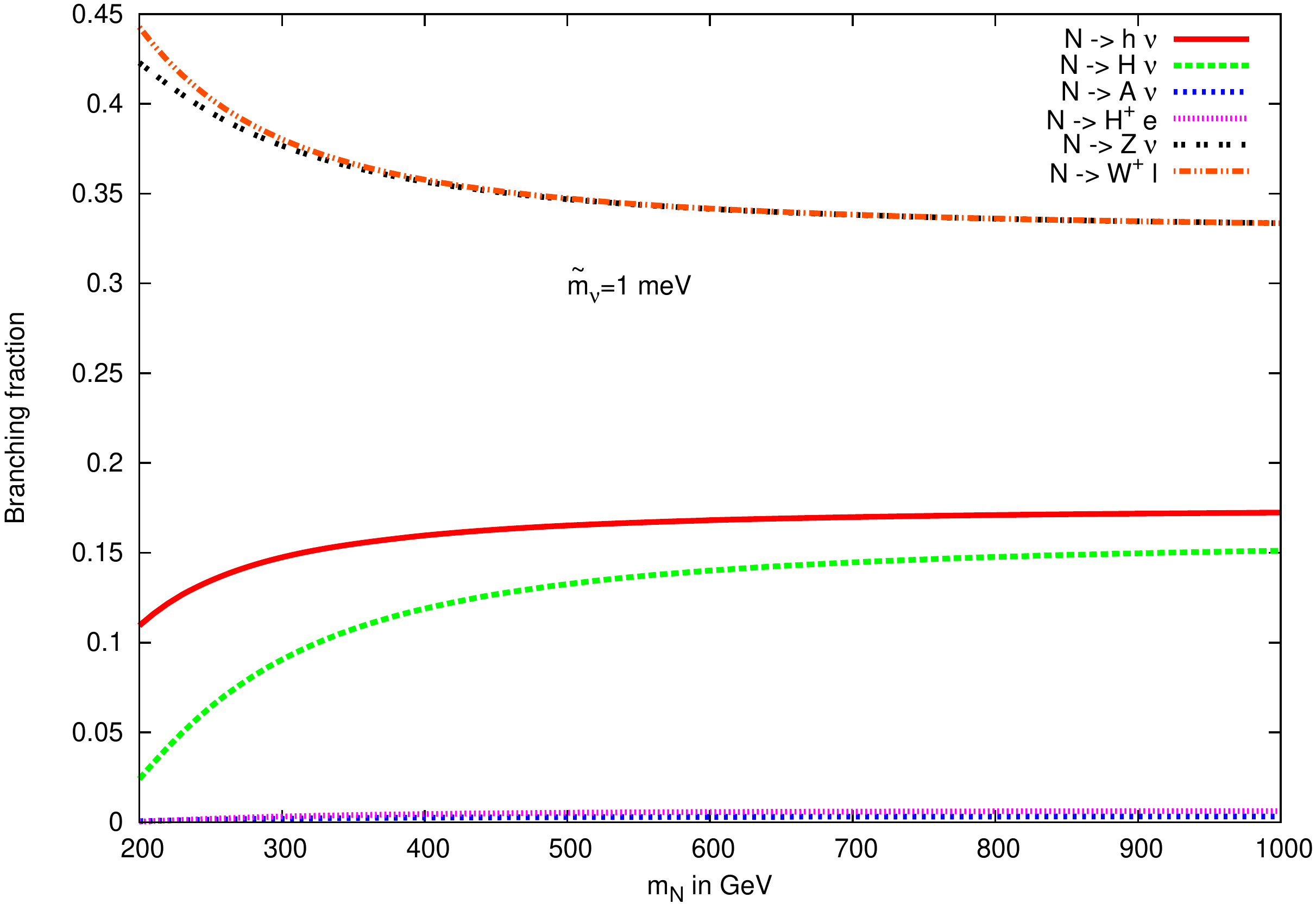}
\includegraphics[width=0.495\linewidth]{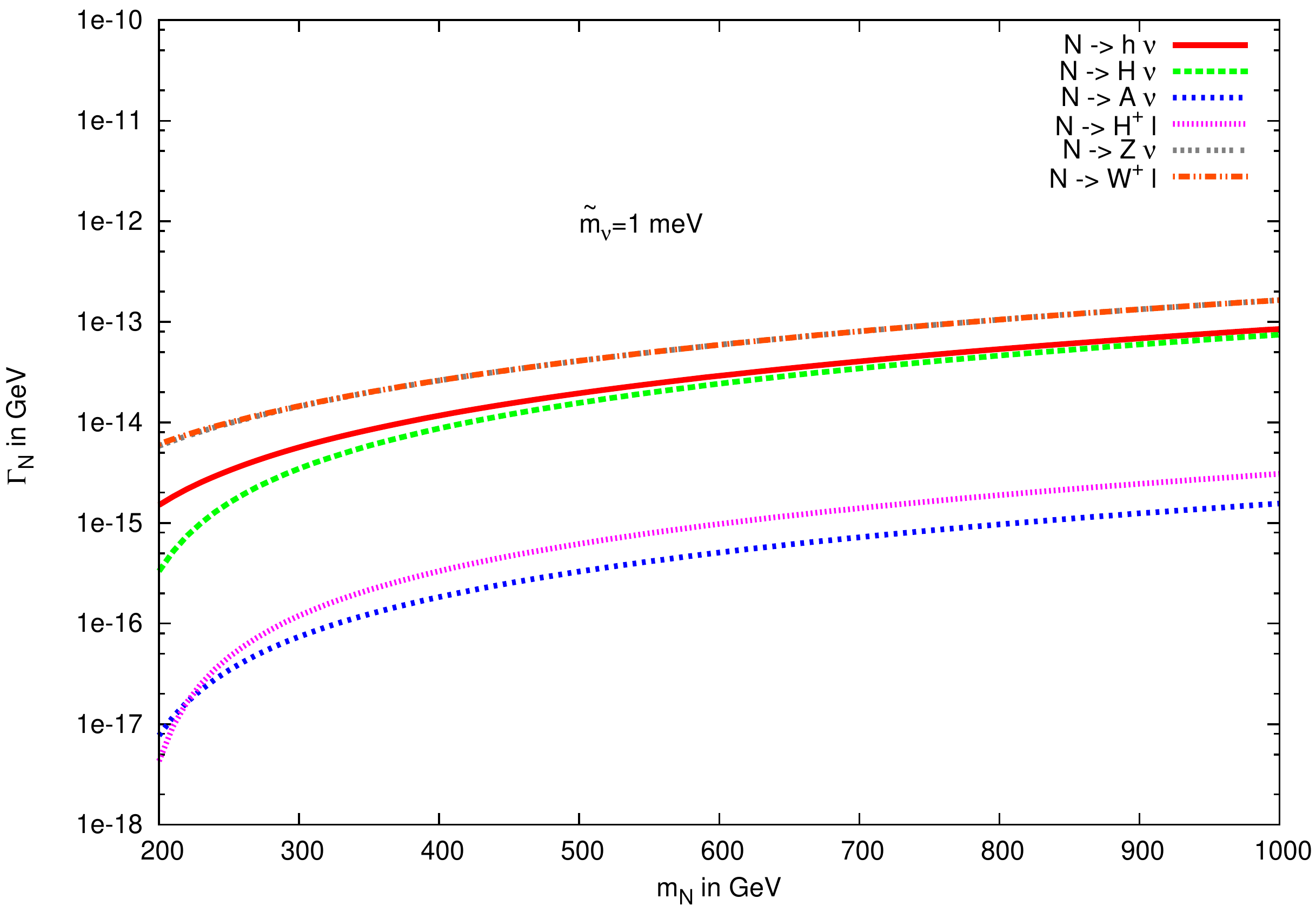}
%
%\hskip -15pt
%{\epsfig{file=plots/sigrn101mev.pdf,width=6.5 cm,height=7.5cm}}
%%\vspace*{-40cm}
%\hskip -12pt
%{\epsfig{file=plots/sigrndcw1mev.pdf,width=6.5cm,height=7.5cm}}
%{\epsfig{file=plots/sigrn100.01mev.ps,width=6.0cm,height=7.0cm,angle=-90}}
%
%\vspace*{-16cm}
\caption{Branching fractions and partial decay widths of the right-handed neutrino, $N$, in the non-decoupling limit of heavy Higgs bosons with $m_A=162.2$ GeV as a function of
$m_{N}$ for the effective neutrino mass, $\widetilde{m}_{\nu}=$ 1 meV and $\tan\beta=10$.}\label{brN}
\end{center}
%\vspace*{-0.5cm}
\end{figure}
%%%%%%%%%%%%%%%%%%%%%%%%%%%%%%%%%%%%%%%%%%%%%%%%%%%%%%%%%

As discussed in the previous subsections, the extra $U(1)_{\chi}$ gauge boson
$Z^\prime$ and
its superpartner $\tilde{Z}^\prime$ can be copiously produced
and decay to right-handed neutrinos at the LHC.
Thus, we can search for the signatures of the pair produced heavy Majorana neutrinos
through two channels:
\begin{eqnarray} \label{NNmode}
&&pp \to  Z'\to N N \,,\\
&&pp \to \tilde Z' \tilde Z' \to  N N \tilde N_1 \tilde N_1\,. \nonumber
\end{eqnarray}
Now, the right-handed neutrino, $N$, can decay to the following final states:
\begin{equation}
N \to l W, \nu Z, \nu h , \nu H, \nu A, l H^+
\end{equation}
as shown in Subsection~\ref{Ndecays}.

Figure \ref{brN} describes the mass variation of the decay
branching fraction of the right-handed neutrino and the
partial decay widths for various decay modes for the
effective neutrino mass $\tilde{m}_{\nu}=1$ meV. From Figure \ref{brN},
it is clear that the right-handed neutrino decays to gauge bosons
and lepton modes have more decay branching fraction $\sim 35-45\%$
for the whole region of the right-handed neutrino mass, $m_N$.
Apart from the production of the gauge bosons from the decay of the right-handed
neutrino, Higgs modes are also possible due to the direct coupling
in the superpotential Eq.~(\ref{wpot}) which is proportional to the
small Yukawa coupling, $y_{\nu}$. The mode $N\to h\nu$ has larger branching fraction
compared to the $N\to H\nu$ mode, which is also expected from
 Eq.~(\ref{ndcy}). The right-handed neutrino decays to
the pseudo-scalar Higgs mode, i.e. $N\to A\nu$, and also the charged Higgs
modes, i.e $N\to H^\pm l^\mp$, have very low branching fraction
 because of the fact that the decay widths are proportional
to $\cos^2{\beta}\approx 10^{-2}$.
For the choice of smaller $\tan{\beta}$, these modes also could be interesting.

In the model under discussion, the Majorana nature of right-handed neutrino,
$N$, can be  probed either of the two channels~(\ref{NNmode}) leading to the 
same-sign dilepton (SSD) final states:
\begin{equation}
 pp \to Z' \; (\tilde Z' \tilde Z') \to l^\pm l^\pm W^\mp W^\mp\;  (+ \ptmiss) \,.
\end{equation}
Depending on the decay of $W^\pm$, the final state can have $3\l$ or
$4\l$, or $\rm{SSD}+4j\, (+\ptmiss)$. The missing energy contribution
 comes from $\tilde{Z}^\prime$ decaying to the LSP, $\tilde{N}_1$.

It is also interesting to look for the Higgs signal from the channel $N\to h \nu$.
When the Yukawa coupling, $y_{\nu}$, or the effective neutrino mass, $\tilde m_\nu$,
is small enough, the Higgs thus produced will be displaced and its main decay to
$b\bar b$ can be observed \cite{Bandyo10}. If the other right-handed neutrino, $N$,
decays to  $l^\mp W^\pm$ or $\nu Z$, then the final states can have one or two 
charged leptons to tag along with the $b\bar{b}$:
\begin{equation} \label{hlW}
 pp \to Z'\,,\; \tilde Z' \tilde Z' \to  h\, l^\pm  W^\mp/Z  + \ptmiss\,.
\end{equation}
To get some reference values for the production cross-section of this signal, let us take $Br(N\to l^\pm W^\mp)\times Br(N\to h\nu) \approx $ 5\% from Figure~\ref{brN} and there is a combinatorial factor 2 as one of the $N$ has to decay to Higgs which makes $Br(NN\to h\nu l^\pm W^\mp)\sim$ 10\%. At the 7 TeV LHC, we have  $\sigma(pp \to Z'\to NN)\simeq$ 0.07 pb for $M_{Z'}=1$ TeV, and $\sigma(pp\to \tilde Z' \tilde Z'\to NN) = 43$ fb for $M_{strong}=1$ TeV assuming the $\tilde Z'$ NLSP as was discussed in the previous subsections. 
This leads to  the production cross-section of the process (\ref{hlW}): $\sigma(h l^\pm W^\mp) =$ 21 fb and 4.3 fb from the $Z'$ and $\tilde Z'$ channel, respectively. Thus, there is a chance to find the Higgs signal at the 7 TeV LHC if, in particular, the associated displaced vertex is large enough to kill the backgrounds.  The corresponding figures at the 14 TeV LHC are $\sigma(h l^\pm W^\mp) =$ 105 fb and 130 fb from the $Z'$ and $\tilde Z'$ channels, respectively.

\begin{figure}[hbt]
\begin{center}
%\vspace*{-4.2cm}
%
%\hskip -15pt
\includegraphics[width=0.70\linewidth]{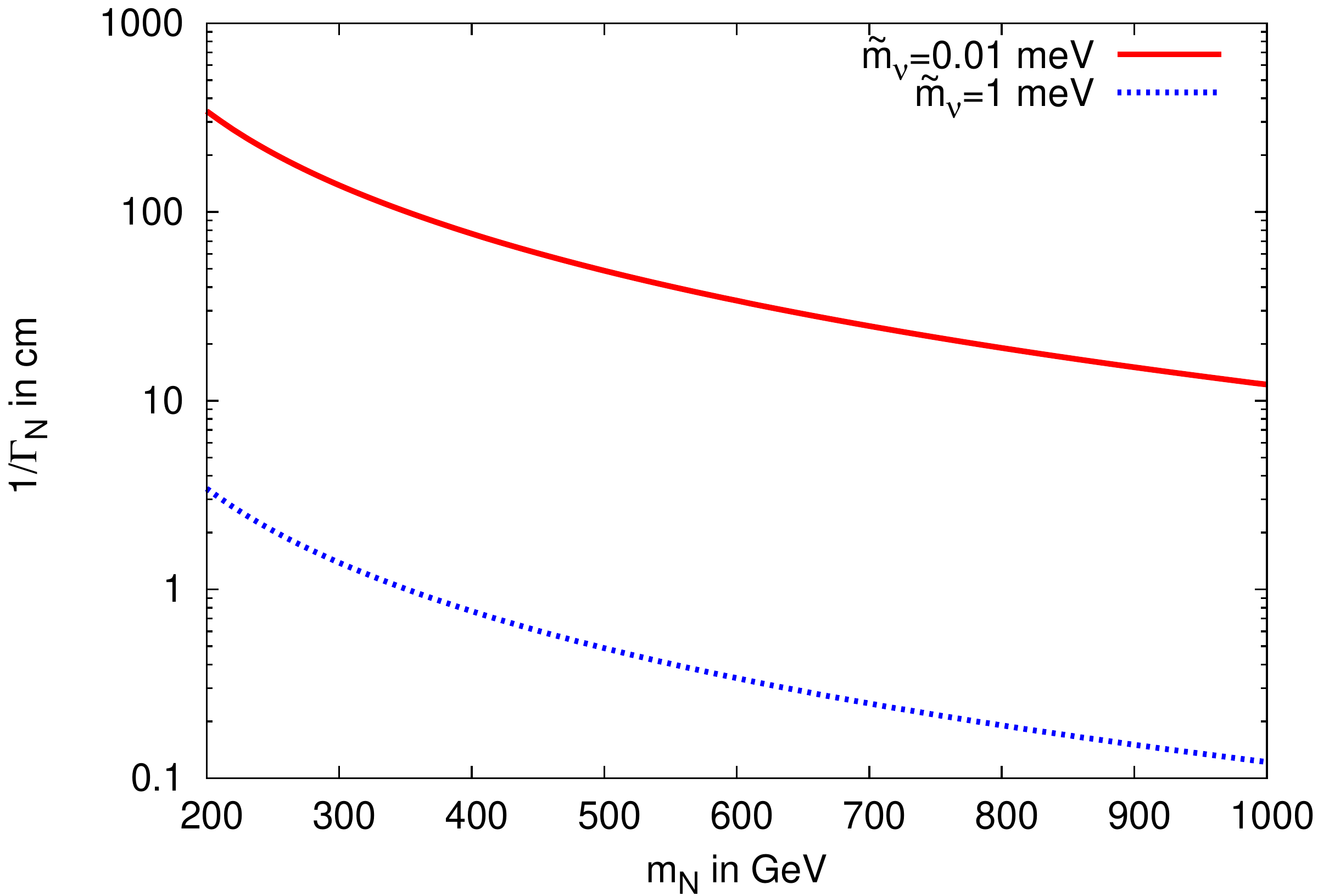}
%
%{\epsfig{file=plots/totdl.pdf,width=8.0 cm,height=10.0cm}}
%
\caption{Variation of the decay length of the right handed
neutrino for $\tilde m_{\nu}=$ 1 and 0.01 meV,
respectively.}\label{tdw}
\end{center}
%\vspace*{-0.5cm}
\end{figure}

Figure \ref{tdw} describes the variation of the decay length of
right-handed neutrino, $N$, with the right-handed neutrino mass, $m_N$,
for two effective masses $\tilde{m}_{\nu}=1$ and 0.01 meV. One can see that
the decay length becomes favorably large for  $\tilde{m}_{\nu} \lesssim 1$ meV
making clean the $N$ decay signals. There is a large parameter space for
such a effective neutrino mass allowing the right-handed sneutrino dark matter
as shown in Figure \ref{fig:mdm}.

%%%%%%%%%%%% Decay lenghts for  N-->2*x  %%%%%%%
%%%%%%%%%%%%%%%%%%%%%%%%%%%%%%%%%%%%%%%%%%%%%%%%%%%%%%%%%

For small $\tan{\beta}$, the right-handed neutrino, $N$, can have a large
branching fraction to the charged Higgs and charged lepton, if kinematically
allowed. Thus, we can have  SSD plus charged Higgs in the final state:
\begin{equation}
 pp \to Z'\; (\tilde Z' \tilde Z') \to  H^\pm W^\pm l^\mp l^\mp\; (+ \ptmiss)
\end{equation}
in which the charged Higgs decays to $\tau\bar{\nu}_{\tau}$ or $t\bar{b}$. These
lead to the displaced multi-jet ($\tau$-jet or $b$-jet) and multi-lepton
final states.

When $\tilde{Z}^\prime$ is not the NLSP,  $\tilde{Z}^\prime$ can also
contribute to $\l \tilde{\l}^*$ if kinematically allowed. In this
situation, we can have the following final states:
\begin{equation}
 pp \to \tilde Z' \tilde Z' \to
 \cases{ W^\pm l^\mp l^\pm l^\mp + \ptmiss \cr
         Z^0 l^\pm l^\mp + \ptmiss \cr
         h l^\pm l^\mp + \ptmiss \cr
         H^\pm l^\mp l^\pm l^\mp + \ptmiss \cr }\,,
\end{equation}
which may involve one displaced and one prompt vertex.

The final states discussed above can be studied
as smoking gun signals for this model at the LHC.
Among those channels, the channels associated with the multi-lepton
final states suppress the SM backgrounds
effectively \cite{Aguila08}.  Again the displaced decay of the right-handed neutrino
actually can remove the SM background completely. Nevertheless, still for the
estimation of signal significance, one needs to consider the
following backgrounds: $t\bar{t}$, $t\bar{t}Z$, $t\bar{t}h$,
 $W+n\rm{-jets}$, and $Z+n\rm{-jets}$ as well as other supersymmetric final states.
% However, these are mainly main
%backgrounds for the case of $2\l$. The final states of $3\l$ and
%$4\l$ are almost background free.
The detailed simulation for the above mentioned final states and the
corresponding significance calculation deserve further investigation~\cite{pbec}.
There is another aspect of this model that could be interesting through
the mixing in the neutrino (\ref{mixingf}) and  sneutrino (\ref{mixing})
sector.Specially, in the context of the NLSP decaying to the LSP ($\tilde{N}_1$),
these can lead to remarkable features in the final state depending on the nature
of the NLSP~\cite{pbec}.

\section{Conclusion}

We considered the possibility of a right-handed sneutrino as the
LSP dark matter in the supersymmetric Standard Model extended to include
an extra $U(1)'$ gauge symmetry realizing the seesaw mechanism with three right-handed neutrinos.
In a supersymmetric seesaw model, a complex right-handed sneutrino gets split into two real mass eigenstates due to the soft supersymmetry breaking Majorana mass term. While
the lightest real right-handed sneutrino, $\tilde N_1$ as the LSP, cannot annihilate through the $U(1)'$ gauge boson  $Z'$ exchange at the s-channel,
its annihilation to lighter right-handed neutrinos, $\tilde N_1 \tilde N_1 \to N N$, mediated by the $U(1)'$ gaugino $\tilde Z'$ at the $t$-channel is shown to be effective in generating the right dark matter relic density.
In this process, the decay and inverse decay of $N$ play
an important role in maintaining $N$ and $\tilde N_1$ longer
in thermal equilibrium and thus reducing the dark matter density.
This behavior was shown in Figure~\ref{fig:xf} by solving the Boltzmann equations with varying the decay rate quantified by the effective neutrino mass, $\tilde m_\nu$.  The resulting dark matter density is computed in Figures~\ref{fig:mdm}, \ref{fig:mnu}, \ref{fig:mZchi}, \ref{fig:g'} showing the favorable parameter space of the model.

The signatures of the model can be probed through the production of $Z'$ as well as $\tilde{Z}^\prime$ and the corresponding decays. In particular, the production of
the right-handed neutrino is of great interest because of its Majorana nature and
possible displaced vertices. When $\tilde Z'$ is light enough as is needed for
the right-handed sneutrino dark matter, it can be pair-produced copiously through
the cascade decays of squarks/gluinos as estimated in Figures \ref{crosstrn}, \ref{brsZ1}
leading to a large number of events for the process of $pp\to \tilde Z' \tilde Z' \to N N + \ptmiss$.  This can be compared with the production cross-section of the usual process $pp\to Z' \to NN$ as shown in Figures \ref{zpprod}, \ref{zpprodss} including all the other final states except the sfermions.  The above channels provide the golden search for the seesaw mechanism and the Majorana nature of neutrinos through the same-sign dilepton final states.
We also point out a remarkable feature of the Higgs production
from the right-handed neutrino decay.
%The signals with $3\l$, $4\l$ and like-sign dileptons are clean
%from the SM backgrounds.
The displaced
$b \bar b$ along with displaced tagged leptons will be a clean signature
in probing the light neutral Higgs. The non-decoupled heavy Higgs bosons can also be probed
in a similar way.  In this article, we have given the number of the production rates
and the effective branching fractions which gives a hint that these
signal topologies can be probed with early data of the LHC for the 7 TeV case.
The detailed collider simulation to calculate the acceptance
under the basics and the hard cuts will be reported in a separate work
\cite{pbec}.

\medskip

{\bf Acknowledgments:} EJC was supported by Korea Neutrino
Research Center through National Research Foundation of Korea
Grant (2009-0083526).

%%%%%%%%%%%%%%%%%%%%%%%%%%%%%%%%%%%%%%%%%%%%%%%%%%%%%%%%%%%%%%

\end{document}